\documentclass[preprint,12pt]{elsarticle}

\usepackage{graphicx}
\usepackage{amssymb}
\usepackage{url}

\usepackage{enumerate}
\usepackage{float}
\usepackage{multirow}
\usepackage{longtable}
\usepackage[utf8]{inputenc}
\usepackage{amsmath,amsfonts,amsthm}
\usepackage[margin=1.0in,a4paper]{geometry}  
\usepackage{algorithmic}
\usepackage{graphicx}
\usepackage{textcomp}
\usepackage{xcolor}
\usepackage{pbox}
\usepackage{natbib}
\usepackage{hyperref}
\usepackage{cleveref}
\usepackage{booktabs}
\usepackage{makecell}
\usepackage{tabularx,booktabs}
\usepackage{diagbox}
\usepackage{enumerate}
\usepackage{graphicx,wrapfig,lipsum}
\usepackage[toc]{appendix}
\usepackage[shortlabels]{enumitem}
\usepackage{acro}
\usepackage{multirow}
\usepackage{colortbl}

\journal{Journal Name}

\begin{document}

\begin{frontmatter}


\title{System Security Assurance: A Systematic Literature Review }


\author[add1]{Ankur Shukla \corref{mycorrespondingauthor}}
\cortext[mycorrespondingauthor]{Corresponding author}
 \ead{shukla395@gmail.com}
 \author[add1]{Basel Katt}
\author[add1]{Livinus Obiora Nweke}
\author[add1]{Prosper Kandabongee Yeng}
\author[add2]{Goitom Kahsay Weldehawaryat} 
 \address[add1]{Department of Information Security and Communication Technology, 
Norwegian University of Science and Technology,  Gjøvik, Norway}

 \address[add2]{Department of Building and Environmental Technology, Norwegian University of Life Sciences (NMBU), Norway}

\begin{abstract}
System security assurance provides the confidence that security features, practices, procedures, and architecture of software systems mediate and enforce the security policy and are resilient against security failure and attacks.  Alongside the significant benefits of security assurance, the evolution of new information and communication technology (ICT) introduces new challenges regarding information protection. Security assurance methods based on the traditional tools, techniques, and procedures may fail to account new challenges due to poor requirement specifications, static nature, and poor development processes. The common criteria (CC) commonly used for security evaluation and certification process also comes with many limitations and challenges.  In this paper, extensive efforts have been made to study the state-of-the-art, limitations and future research directions for security assurance of the ICT and cyber-physical systems (CPS) in a wide range of domains. We conducted a systematic review of requirements, processes, and activities involved in system security assurance including security requirements, security metrics, system and environments and assurance methods. We highlighted the challenges and gaps that have been identified by the existing literature related to system security assurance and corresponding solutions. Finally, we discussed the limitations of the present methods and future research directions.

\end{abstract}


\end{frontmatter}
\section{Introduction}
Recent advancements in information and communication technologies have revolutionized the entire social and economic systems. In this information age era, government and commercial organizations heavily rely on information to conduct different activities.  Alongside significant benefits, the ever-increasing criticality, connectivity, and comprehensiveness of software-intensive systems introduce new challenges for cybersecurity professionals to protect the information.  Compromise in confidentiality, integrity, availability, accountability, and authenticity of information and services can harm the operation of the organizations, and it is needed to protect the data and information of IT systems within the organization. Therefore, it has become a crucial task for security researchers and practitioners to manage the security risks by mitigating the potential vulnerabilities and threats with the help of new techniques and methodology and achieve the acceptable security assurance of an IT system so that the stakeholders can get greater confidence that the system is performing in intended or claimed way with acceptable risks. 

 Several definitions of security assurance can be found; however, the common basis of these definitions refers to the trust and confidence in the secure and correct operation of (software) systems. NIST defined security assurance as a “measure of confidence that the security features, practices, procedures, and architecture of an information system accurately mediates and enforces the security policy” \cite{radack2011managing}. Katt and Prasher \cite{katt2018quantitative} defined security assurance as “the confidence that a system meets its security requirements and is resilient against security vulnerabilities and failures.” They further defined confidence as the level of trust of a system that is safe to use. However, there is a difference between the “security need” and “security assurance”. Jelen and Williams \cite{jelen1998practical} pointed out this fact and defined security needs as a threshold value on that the measurement of actual level can be made. The security level can be measured by comparing the measured value and the threshold value; however, the confidence of the calculation depends on the accuracy of the measurement. Considering these aspects, they defined assurance as “a measure of confidence in the accuracy of a risk or security measurement”.   
   
Security assurance has always been the keen interest of researchers and practitioners. Security assurance tries to address two essential questions ``Does a system do what it is supposed to do" and ``Does the system do anything unintended?" \cite{spafford2003testimony}. Security assurance activities go throughout the development life cycle of the software system from initiation of the protection profile to certification of a target of evaluation (TOE), which is the system that will be assessed and evaluated.  Security assurance is also continued when the system is in the operational phase. Different scanning tools can be used to ensure and maintain the continued security by locating and patching security errors and vulnerabilities.  Security testing and evaluation are beneficial to get the desired level of security assurance. However, there is no single standard process available to measure the security assurance of the software system. One can get certain degree of confidence regarding the system's security and its components by reviewing its development process. The level of confidence can be increased if a rigorous methodology of security requirement definition, design, specification, and conformance have been considered. The past users' experience on a particular system can also provide some degree of assurance. On the other hand, if multiple organizations use a system without any security incidents, one can trust that it will operate securely in their organization.    Application of some new technologies such as advanced software engineering also provides assurance. 
   
In the past, various standards and frameworks have been developed to evaluate the security assurance of the system. The initial effort was made by Trusted Computer System Evaluation Criteria, a United States Government Department of Defense standard, which is also known as the Orange Book, to assess the system's security. Some other efforts have been made in this continuation, such as European developed Information Technology Security Evaluation Criteria(ITSEC), Canadian Trusted Computer Product Evaluation Criteria (CTCPEC), ISO SC27 WG3 security evaluation criteria, etc. Later, these criteria are primarily integrated into single criteria, i.e., CC. OWASP Application Security Verification Standard (ASVS) is an open standard that can be used for technical security control testing of web applications. OWASP also provides the requirements list to developers for secure development.   Software security maturity models lay out the requirements of different security levels and software engineering and maintenance practices that fulfil those requirements. Some examples of software security maturity models are  Building Security In MaturityModel (BSIMM), BSIMM for vendors (vBSIMM), and OWASP’s Software Assurance Maturity Model (OpenSAMM) \cite{heiland2013toward, stock2019application}.
   
In the past, several efforts have been made for system security assurance and its evaluation. The major focuses of these researches are to provide solutions to ensure the security of the systems used in various application domains and environments by developing security assurance methods and techniques. These solutions include different methods, techniques, processes, and recommendations such as operational security assurance, security assurance requirements engineering methodology, security assurance metric and aggregation techniques, etc. These works also include developing security assurance methods for composed systems that are made with different components such as protocols, servers, clients and services. Some efforts have also been made towards early detection of security vulnerabilities, development of a security assurance model, and security assurance tools to maintain and enhance the security of the deployed system.  Security assurance is also essential for the software developers to address the security concerns in the early development and acquisition phase, and to measure their preparation towards the advancement of secure software. The development of secure software requires considerations beyond the basic security requirements  such as authentication/authorization and mandated operational compliance to identify and resolve the risk environment in which the system must operate. Some authors have considered the security assurance methodology in the different development life cycles of the software. However, these methods and techniques come with several drawbacks.  The main drawbacks of these approaches are that they are static, time-consuming, and do not scale well to the extensive, networked, IT-driven system.  It also does not offer continuous security assurance. Many researchers have made efforts to resolve these challenges. However, it is still an open issue.   

In the past, no significant efforts have been made on systematic literature review (SLR)  on system security assurance. Some studies can be found; however, they focused on a particular security concern of a specific application or application domain. Therefore, one cannot get a clear and comprehensive overview of the existing security assurance approaches, related information, and evidence.   Therefore, a detailed and systematic literature review on “System Security Assurance” has been conducted in this paper.   The motive of this paper is to study state-of-the-art, research trends, limitations, and future research directions in security assurance of the ICT and CPSs in a wide range of domains. It will also investigate the conventional and emerging technology for security solutions. This paper provides detailed information and discussion on the research challenges and gaps, the efforts made toward these challenges and gaps, limitations of these approaches, and future research directions. 
   
The rest of the paper is organized as follows: In Section 2, the existing works related to the security assurance survey have been discussed and the need for the survey has been established. Section 3 presents a detailed discussion of the methodology of the SLR. In section 4, 5 and 6, the detailed discussion about security assurance process, the role of CC, and challenges and gaps in the existing methods and technology have been discussed respectively.  Security requirements and security metrics is discussed in section 7. Section 8 presents the different security assurance methods. Security assurance methods developed and applied on different systems and environments is given in section 9.  In section 10, limitations and future directions is discussed. Finally, section 11 concludes this paper.

\section{Related Works and the Need of This Survey}

In the past, very limited surveys have been published related to system security assurance. However, no dedicated works that consider detailed and systematic study covering processes and activities involved in system security assurances could be found. In this section, the existing surveys have been summarized and compared with our work. As observed in the literature, most of the works do not follow the systematic methodology in conducting the literature review. These works are either focused on a specific application domain or considered only limited aspects of the security assurance process. Summary of the topic covered in the existing literature, their contributions and limitations are given in Table 1.  The enhancement made in our work is also given in this table. 

  Choi and Yoo \cite{choi2009software} conducted a study on software assurance and discussed the critical security flaws and vulnerabilities related to software installation and software execution. They proposed a system for software assurance. This study incorporates some issues related to software security. This study considered the researches carried by limited government agencies and research institutes. 
   
\begin{table}[!h]
\caption{Details of related works, their limitations, and enhancement in  this paper.}
\scriptsize
\centering
\begin{tabular}{p{.03\textwidth}p{.03\textwidth}p{.16\textwidth}p{.21\textwidth}p{.20\textwidth}p{.22\textwidth}}
\hline
{Year} & {Paper} & {Topic(s)}& {Contributions }& {Limitation }& {Enhancements in our paper }\\ \hline
2009& \cite{choi2009software} & Software assurance & Investigated research on software assurance and proposed a software assurance system  & Considered security flaw related to software installation and vulnerability related to software execution only. & Focuses on general security assurance and considered security requirements and vulnerability in a wide range of application domains. \\
\hline
2013&\cite{brown2013toward}& Communication security using formal models &A survey of formal models of communications security and taxonomy of security concerns.  & Focused on communication security using formal models. &Considers on the overall system security assurance.
\\ \hline
2013&\cite{guo2013security}& Security-related behaviour &A review on security-related behaviour in the workplace and a framework for conceptualizing security-related behaviour.  &Mainly focused on security-related behaviour. &Considers broader aspect of security assurance
\\ \hline
2014& \cite{bijani2014review}&Security of Open multi-agent systems &A survey on security techniques for multi-agent systems. &Focused on multi-agent systems and do not consider security assurance methods. &Considers security assurance for wide range of application domain. 
\\ \hline
2014& \cite{wan2014context}&Context-aware security solutions for CPSs &A survey on the state-of-the-art of CPSs to identify the security issues and an investigation the role of context-awareness to improve the extent of CPS security. &This survey is very limited and mainly focused on context-aware security solutions.&Extensive review on security assurance of CPS.
\\ \hline
2015& \cite{oueslati2015literature}& Development of secure software using agile approach &Literature review of the challenges in the development of security software using agile approaches. &This work is very limited and mainly focused on the agile method in the development of secure software. & Extensive review which is not method based and considers every software development life cycles.
\\ \hline
2015& \cite{zhang2015survey}&Security of information systems &A survey on security of information systems &This work do not consider security assurance and evaluation processes.& Presents a detailed overview of security assurance and evaluation in a wide range of domains.  
\\ \hline
\end{tabular}
\end{table}

        Brown \cite{brown2013toward} surveyed the various hierarchically ordered and adjacent sciences, notations, and security requirements analyses which are essential for extensive communication security.   They developed a taxonomy that provides a comprehensive framework for identifying and analyzing security requirements and potential attacks. This study is focused on formal methods in securing the system.
     
        Guo \cite{guo2013security} discussed the security-related behaviour in the workplace. They reviewed different concepts of security-related behaviour and developed a framework for conceptualizing security-related behaviour to delineate and synthesize the difference between the divergent concepts. This research work is focused on security-related behaviour and does not consider other security perspectives. 
     
      Bijani and Robertson \cite{bijani2014review} conducted a review of security techniques in literature and suggested the appropriate security technique for a class of attacks in open multi-agent systems. Wan and Alagar \cite{wan2014context} studied the state-of-the-art of security of CPSs. The focus of this study is to provide context-aware security solutions for CPSs. Ouchani and Debbabi Ouchani and Debbabi \cite{ouchani2015specification} conducted a review to study the state-of-the-art of security requirements specification, attack modelling, security requirements verification, and security quantification for the software and systems that are based on Unified Modeling Language (UML) or Systems Modeling Language (SysML).    Oueslati et al. \cite{oueslati2015literature} conducted the literature review to identify the challenges and issues in the development of secure software using the agile approach.     Zhang et al. \cite{zhang2015survey} surveyed cybersecurity. They discussed the research and development in cybersecurity.

     The current literature covered various security issues and challenges in different domains and investigated the conventional and emerging technology for security solutions. However, the survey which is focused on system security assurance is still missing. On the other hand, most of the present works do not follow the systematic process in the literature review. Due to this, the existing reviews fails to deliver a clear and comprehensive overview of the available information and evidence on system security assurance. Also, they fail to identify the actual research challenges and gaps in this field. We conducted a systematic and extensive review of system security assurance in this paper to overcome this situation. 

\section{Methodology of SLR}
 
A SLR provides a systematic, explicit, and reproducible way to identify, select, evaluate, and critically appraises the existing body of completed and recorded research works \cite{fink2019conducting}.  The main motive of this SLR is to evaluate and interpret the recent research on system security assurance to address the current research problems and challenges. To conduct the systematic and fair evaluation of literature, a review protocol has been established. The construction of guidelines for this SLR is derived from the “Guidelines for performing Systematic Literature Reviews in Software Engineering” \cite{keele2007guidelines}. Some other guidelines for systematic review have also been reviewed \cite{kitchenham2007cross} to developed the review protocol. 
 
The guidelines established for this SLR include mainly three steps: review planning, conducting the review, and review report.
  \subsection{Review Planning}
   In this stage of the SLR, the following points have been addressed: 
\subsubsection{Purpose of SLR}
    The main purpose of this literature review is to study the current challenges and gaps in system security assurance.  This SLR will conduct a detailed study of system security assurance requirements, metrics,  frameworks, and methods.  The security assurance of different systems and environments in a wide range of applications domains will be discussed.  Specifically, the purpose of this SLR can be summarized as follows: (a) to study essential background, state-of-the art, research trends and directions in system security assurance, and (b) to develop a taxonomy on system security assurance.

\subsubsection{	Developing review protocol }

 A review protocol is developed with the detailed review of the existing methodology of SLR and discussion with the experts.  This protocol includes the design of research questions, search strategy, and potential resources. Furthermore, study selection criteria, selection procedure, and quality assessment checklists have been described in this protocol.  In the protocol, the data extraction strategy and synthesis of the extracted data have been specified.  The protocol is also focused on some other planning and management information such as dissemination strategy and project timeline.       
\subsubsection{Review protocol evaluation }
The review protocol has been reviewed rigorously against the following criteria:
\begin{enumerate} [(i)]
    \item  Whether search strings are appropriate and match with the research questions?
    \item  Whether points of the data extraction will address the research questions properly? and
    \item  Whether analysis procedure is appropriate to fulfil the objectives of the SLR?
\end{enumerate}

\subsubsection{	Specifying  research questions} 
    
       The following research questions have been considered of this SLR based on reflection, debate, and reformulation: 
    \begin{enumerate} [start=1,label={\bfseries RQ\arabic*.}]
\item 	What are the current trends and results related to security assurance considering process, methods, guidelines, tools, metrics, evaluation/techniques, automation, standards, and application domains?  
\item 	What are the challenges, limitations, and gaps related to security assurance? 
\item 	What are the future directions/trends related to security assurance?
\item 	How can we categorize/classify the different research activities related to security assurance?
\end{enumerate}


\subsection{ Conducting the review }
After the proper establishment of the review protocol, the review process  started. 
\subsubsection{Identification of Research}
Identification of primary studies related to the research questions using an unbiased search strategy is an important step. Initially, some search strings are derived from the research questions. These search strings are constructed using Boolean ANDs and ORs. These search strings are tested against the existing primary studies on system security assurance from well know databases. Based on the testing results and discussion with the expert, two search strings are finalized, which are: (i) 
 “Security Assurance”, and (ii)  System and Security and Assurance.

Thereafter, an exhaustive search has been performed considering the six electronic sources including:	IEEExplore, 
	ACM Digital library,  
 	Google Scholar, 
 	Science Direct 
	SpringerLink, and
	Wiley online library.

\subsubsection{	Study Selection}
After identifying the potential relevant primary studies, they need to be assessed based on their relevance. Selection criteria are helpful in identifying and selecting primary research studies that provide evidence for the research questions.  Therefore, selection criteria have been decided based on the research questions to reduce the likelihood of bias. In this SLR, the following inclusion criteria have been used:
\begin{enumerate}[(a)]
\item Papers that were published between 2004-2020.  
\item Papers that focus on the security assurance assessment and evaluation of the systems, for example, security requirement analysis, security assurance framework, and security models.
\item 	Security assurance papers that focus solely on ICT and cyber-physical systems.
\end{enumerate}
The exclusion criterion is “papers that are not related to information and communication technology or cyber-physical system”. The selected papers were also updated continuously, and the last update was made in March 2021. We have selected the starting year 2004 because we aim to study the recent developments in system security assurance over the last one and half decades.   
  
\subsubsection{Reliability of inclusion decisions }
To improve the reliability of the inclusion decisions, each electronic source is assigned to two researchers. Accordingly, each paper is assessed by two researchers based on the inclusion and exclusion criteria. To measure the agreement between researchers, a list of included/excluded papers with the reason for inclusion/exclusion has been maintained. In case of disagreement or misunderstanding on any paper, it has been discussed in the common meeting of the group and resolved by discussion and with expert advice.

\subsubsection {Study Quality Assessment }
After collecting all potentially eligible articles, the next step is the quality assessment to examine the articles more closely. The primary purpose of the quality appraisal is to conduct the second screening to eliminate the articles that are not relevant to this study. As a result, in addition to general inclusion and exclusion criteria, it is essential to examine the quality of primary studies. Considering this fact, now stricter criteria have been established. The followings are the quality assessment criteria: (a) are the aims clearly stated? (b) is the research method used appropriately? (c) does the research work evaluate the outcome appropriately? and (d) does the research work allow the questions to be answered?
      \begin{figure*}[h]
        \centering
        \includegraphics[scale=0.48]{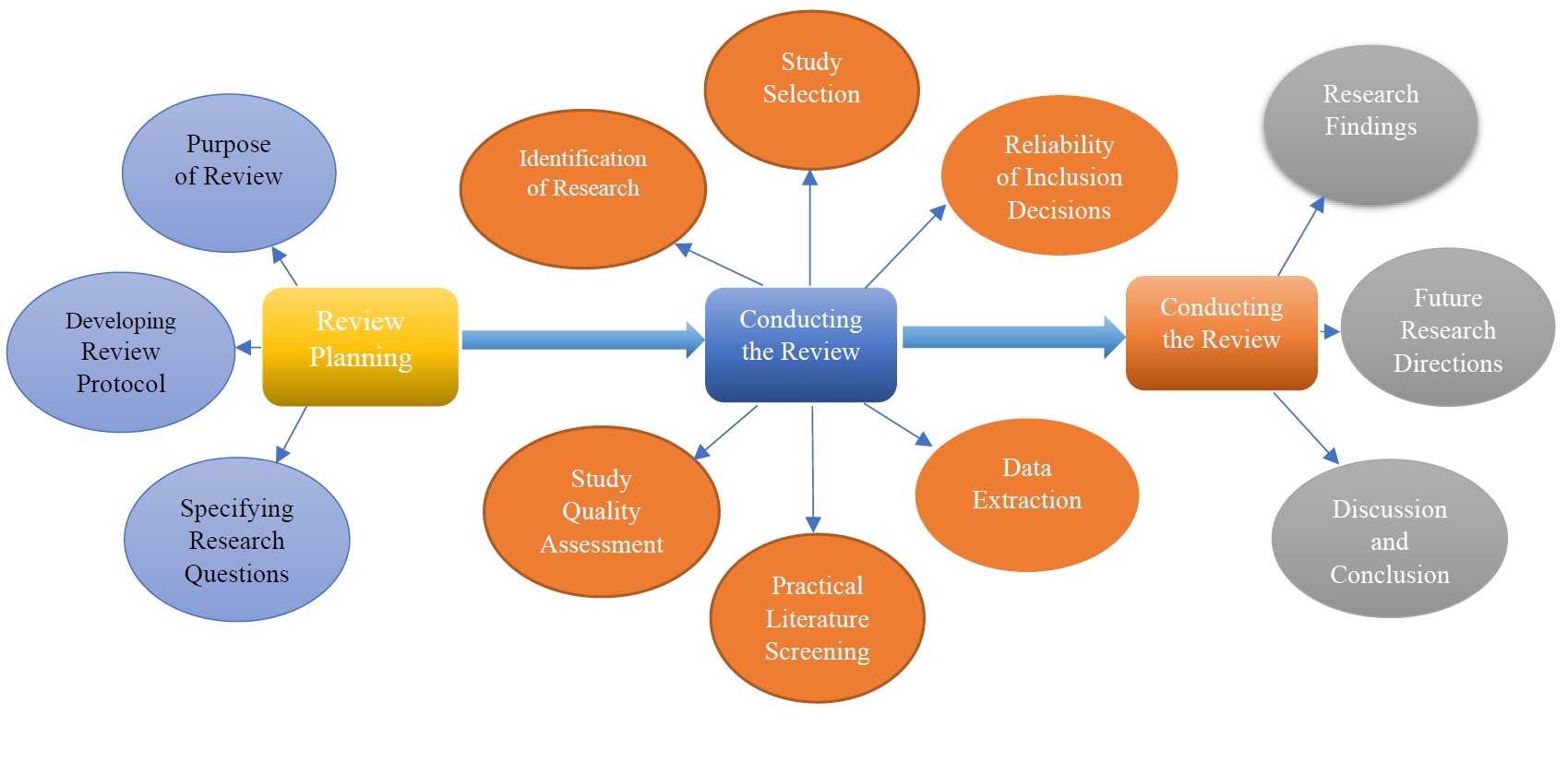}
        \caption{Methodology of SLR.}
        \label{fig:my_label}
    \end{figure*}

\subsubsection {Practical Literature Screening}
Considering the above steps and criteria, the literature screening has been conducted. Results of the different rounds are as follows:
\begin{enumerate}[(a)]
    \item Round 1: The literature is searched and collected from different electronic sources. In this round total of 733  literature items were collected. The duplicate entries were also eliminated and after elimination, the remaining number of literature items are 564. 
    \item Round 2: In this round, quality assessment of the literature was conducted using the aforementioned quality assessment criteria, resulting in a total of 90 literature items. \end{enumerate}

\subsubsection{Data Extraction }
The data extraction form has been designed to collect helpful information to answer the research questions. The data extraction form has been piloted on a sample of a preliminary study to assess the form's completeness and avoid any technical issues. The different points of the data extraction form have been decided based on the research questions, and each point has been defined clearly to avoid misunderstanding, misinterpretation between the researchers. The key points of the data extraction form and their definitions are given in APPENDIX A.  

\subsection{Review Report}
The report or writing the review is the final step of developing a research literature review. The process includes reporting and writing the findings systematically and smoothly so that the entire process can be reproducible scientifically. A pictorial illustration  of the SLR methodology of this survey is shown in Fig. 1.

\section{Security Assurance: Definition, Process, and Types}

We define system security assurance as \emph{the confidence that a system meets its security requirements and is resilient against security vulnerabilities and failures}.  Security assurance is a complex and time-consuming process that goes throughout the development life cycle of a (software) system begins from the protection profile initiation to the TOE certification.  The security assurance process of a system requires a set of inputs such as TOE, the operational environment, assessment criteria and requirements (assurance profile), assurance methods, and assurance level \cite{katt2018quantitative}.  This process goes through multiple stages and involves various activities such as defining security goals, security requirement analysis, threat analysis, vulnerability analysis, penetration testing, security audit, scoring, and analysis, etc. The output of the security assurance process provides the security assurance level and other useful information, recommendation, and mitigation plan that help stakeholders to improve confidence, align with best practices, and reduce the risk following a cyber-attack.  The security assurance process and its essential components are presented in Fig. 2. 
    
\begin{figure}
        \centering
        \includegraphics[scale=0.20]{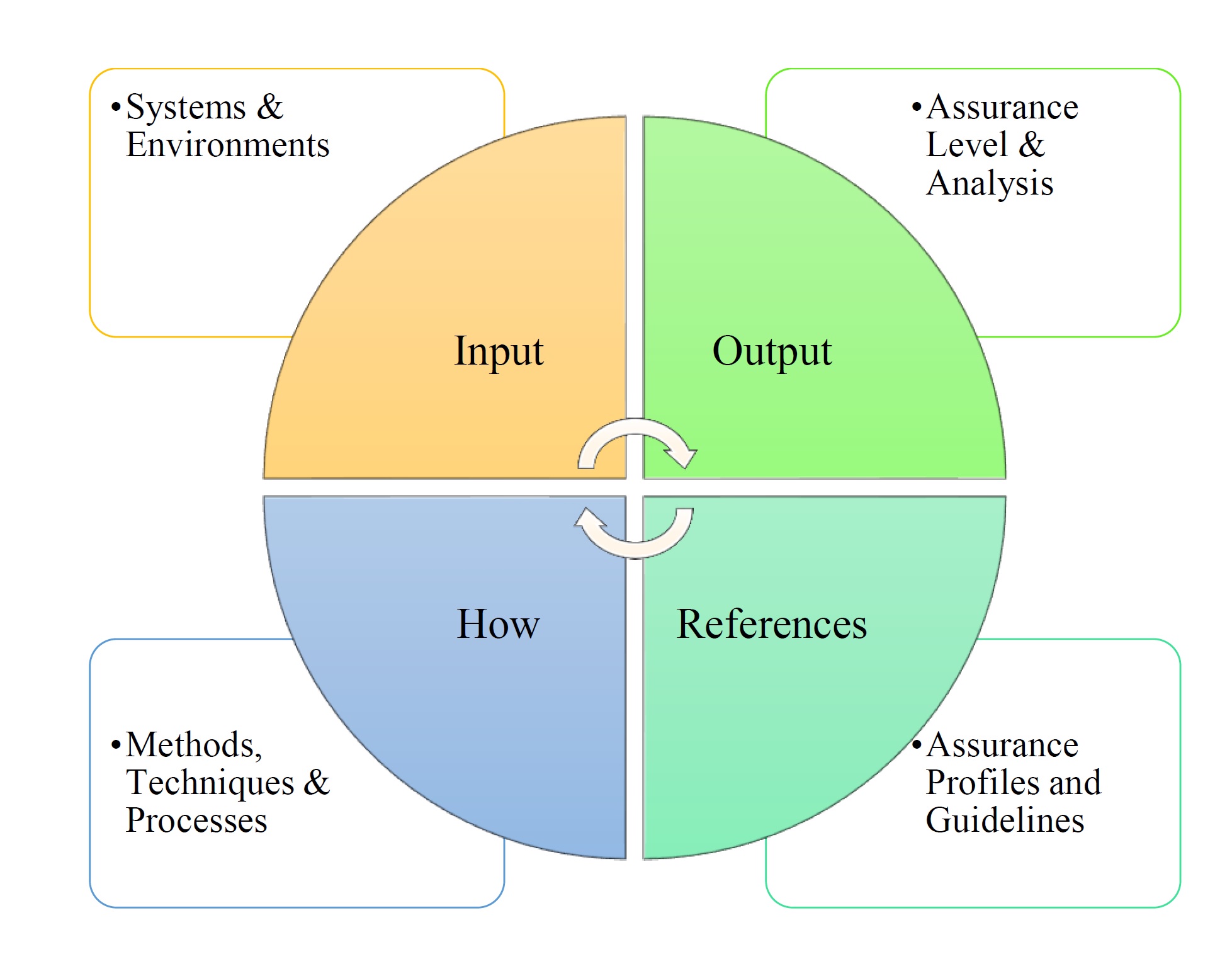}
        \caption{Security assurance process.}
        \label{fig:my_label}
   \end{figure}

\subsection*{ (a) Operational Security Assurance}
    Assurance activities are  crucial during the operations and maintenance phase to ensure that the assurance level of a system to which it is certified, is maintained.  The security requirements specified for a system may be violated in the operational phase because of improper implementation of the security measures, hazardous environment, or invalidity of the assumptions under which the security requirements were specified \cite{ouedraogo2010information}. In other words, the security requirements identified during the development phase based on the assumption made on the system's operational environment may no longer be valid if there are any changes in the system environment. Therefore, it is required to collect evidence to verify the fulfilment of security requirements of the system in the operational phase\cite{ouedraogo2010agent}.  Therefore, one should also focus on security assurance after the deployment or implementation phase. The security assurance evaluation of the system in operation comes with many challenges as well as benefits that cannot be accomplished by an offline assessment. The operation security assurance is complex due to the openness, aggregation, and dynamics nature of IT and cyber physical systems \cite{hecker2009operational}.
    
    \subsection*{(b) Continuous Security Assurance}
Organizations are struggling to ensure security a routine element of their operations. They are exposed to a number of risks that necessitate the deployment of compliance and security controls. Continuous security assurance can be a potential way to manage these risks with continuous monitoring, continuous compliance, and continuous security \cite{ouedraogo2008deployment, ardagna2018case}. On the basis of evidence collecting, continuous security assurance also reports if the security requirements are met throughout system operation \cite{ouedraogo2010agent}. 
        
    \subsection*{(c) Optimal Security Assurance}
It is not possible to make the software systems completely secure. Some vulnerabilities may be present, which were not fixed during the development process due to time constraints or other reasons, and these must be re-examined, prioritized, and fixed. Security assurance is a very time-consuming and costly process. Optimal security security assurance aims to provide optimal security and to reduce these costs\cite{kumar2020knowledge}. 

 \begin{figure}
        \centering
        \includegraphics[scale=0.08]{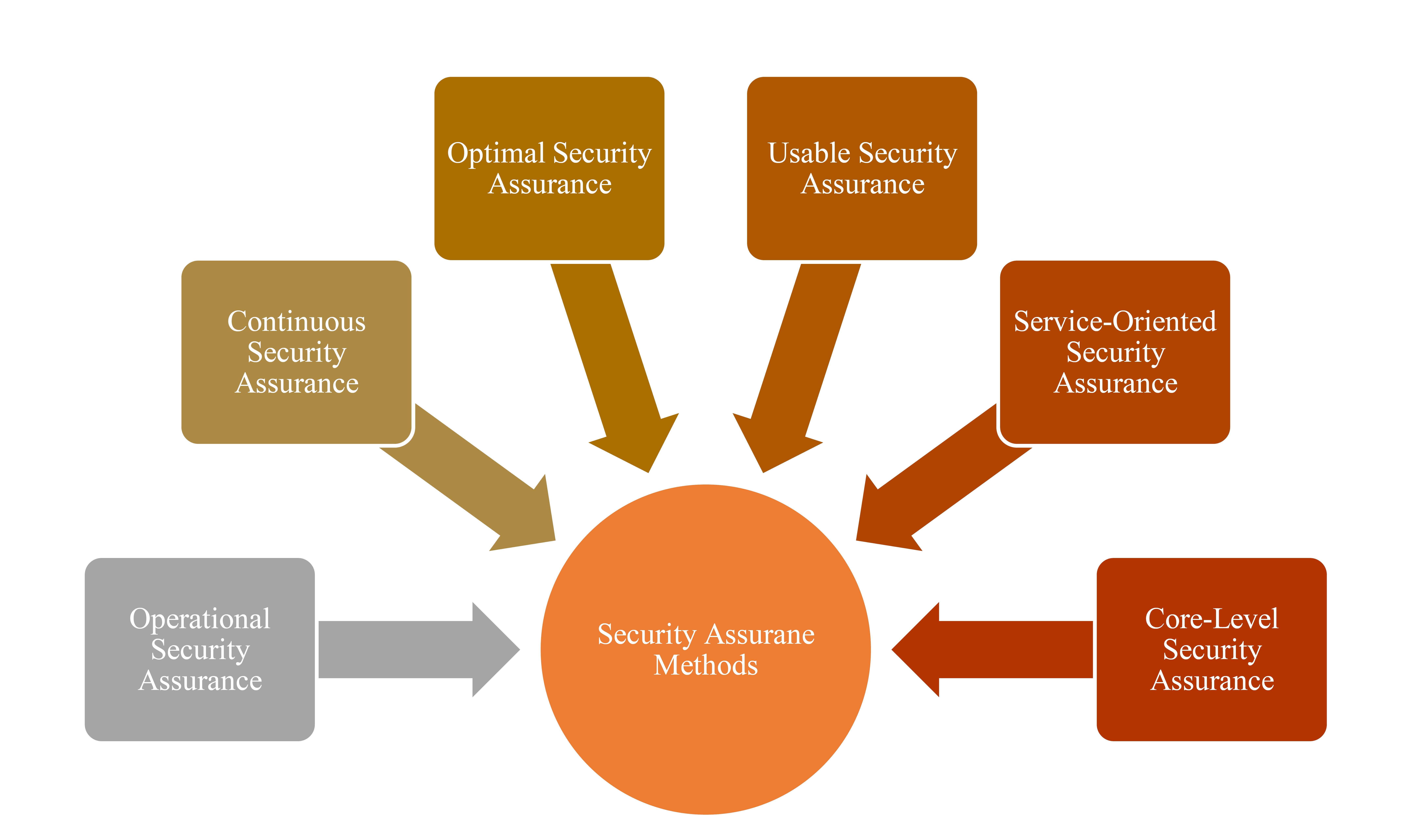}
        \caption{Different types of security assurance.}
        \label{fig:my_label}
     \end{figure}
    \subsection*{(d) Usable-Security Assurance}
   The security goals are mainly focused on the user’s demand, and demands are changed when there is a change in their requirements. The security goals can be achieved through the rigorous testing, establishment, and assessment to provide the defense against malicious attacks. However, the system user can sometimes be the weakest link and may unintentionally invite attacks. Therefore, it is vital to secure the system from the threat as well as maintain usability. The usability focus on the ease of users ‘keeping simple’ formula \cite{liu2011analyzing, tilson1998factors}.  

   \subsection*{(e) Service-oriented Security Assurance}
    The business decomposition process into services is a possible way to provide the flexibility to adapt to the changes in the business needs of the enterprises. This can be provided by a service-oriented architecture, which allows the user to find and use services dynamically. On the other hand, security in service selection is also a crucial factor. Therefore, service-oriented assurance is required to evaluate the security of sub-services. Data from various sources are required to assess the security properties in this process. In addition, the system states (such as established security policies), events, certificates, and other security verification evidence from the third parties are required. The study proved that security properties could be specified and verified objectively in various services using the security-oriented assurance model \cite{karjoth2005service}.

 \subsection*{(f) Core-level Security Assurance}
The operating system core can be considered instead of the application level service in order to increase the speed and effectiveness of attack detection. The main advantage of considering the operating system's core is that it contains every internal attribute and the file system \cite{sakthivel2020core}.

\section{Common Criteria for Security Assurance} 
        The meaning of security may vary from person to person and from organization to organization.  Therefore, common security standards are essential for IT systems with complex and diverse configurations. In this regard, the need for CC was realized to evaluate the security of an IT product. The origins of the CC are discussed in the Introduction section. The CC for information technology security evaluation is a well-known international standard (ISO/IEC 15408) for computer security certification\footnote{https://www.commoncriteriaportal.org/cc/}. The current version of CC is 3.1 revision 5. It provides a set of guidelines and specifications that can facilitate the specification of security functional requirements and security assurance requirements. While the security functional requirements define the expected security behavior of information security products and systems, the security assurance requirements demonstrate that the security attributes have been implemented correctly. CC evaluation framework is given in Fig. 4.  
         
\begin{figure}
        \centering
        \includegraphics[scale=0.10]{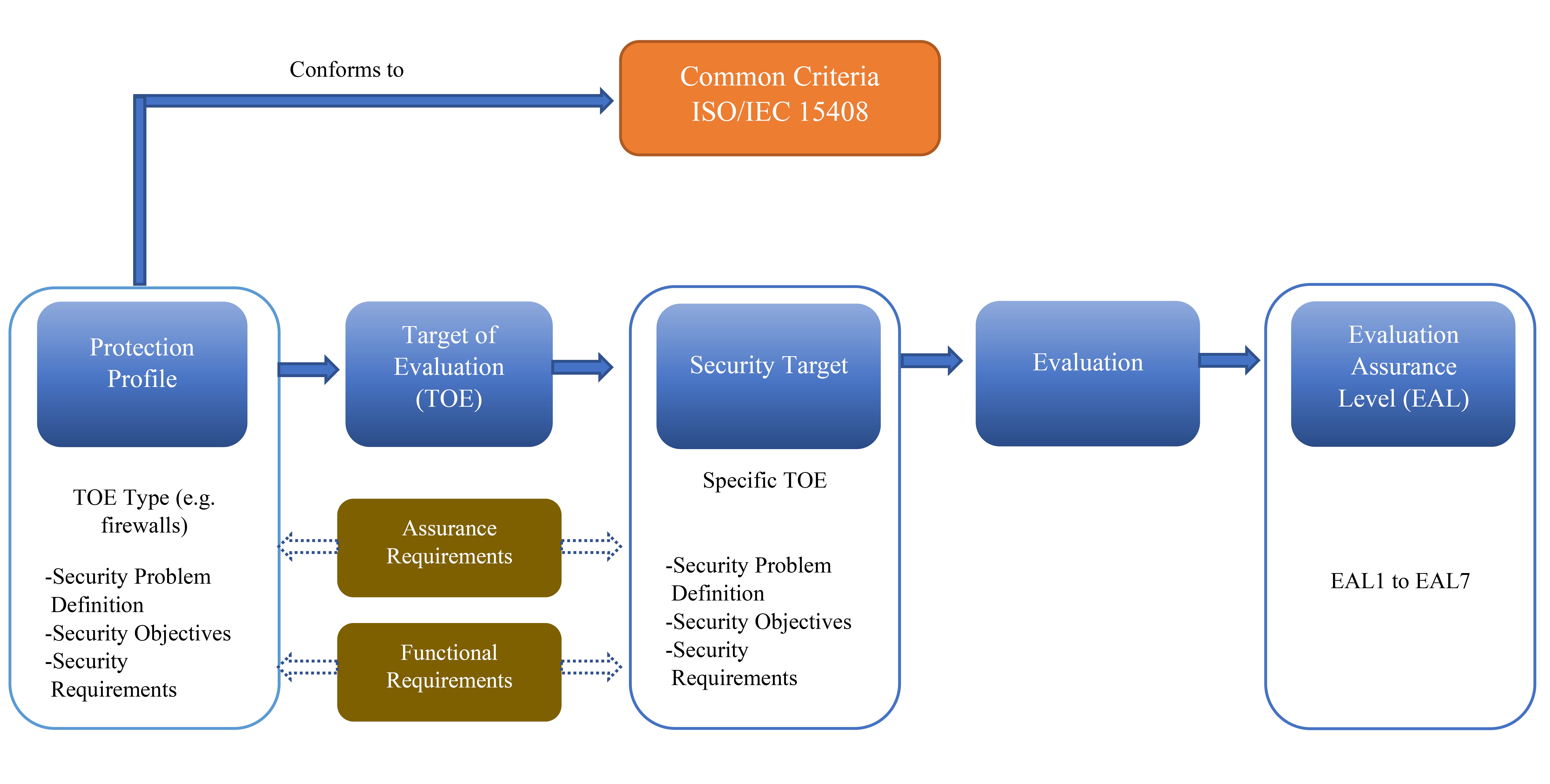}
        \caption{CC evaluation framework.}
        \label{fig:my_label}
   \end{figure}
         
         The general objective of CC is to provide a framework that allows users to specify security functional requirements, enable the developers to specify the security attributes, and help evaluators to ascertain if the security attributes as defined by the developers meet their claims.      The use of CC in the development of information security products and systems can improve the overall security of the products and reduce the time and cost of IT security evaluation. For example, Kim and Leem \cite{kim2004case} showed a method that employs CC in the development process to improve the security of software products. They used a case study that involves the development of MTOS7.5, a security-enhanced UNIX-like operating system based on BSD 4.4 according to EAL3 in CC. The results from their study indicated that CC applied to the development process of software products can enhance the security of the products, reduce the time and efforts in developing the products, and shorten the evaluation periods of the products.


     CC can also be used at the early stages of the software lifecycle to integrate requirement engineering and security engineering to develop secured information systems. This approach is proposed by Mellado et al. \cite{mellado2007common} and involves the use of a CC-centered and reuse-based process to address security requirements at the early stages of software development. The authors utilized a security repository that they combined with CC and then applied to the early stages of the software life cycle. The objective was to merge the ideas of requirement engineering and security engineering.
     
     In the literature, other works can be found that have used CC in the development process of information security products, requirement engineering, security engineering, and certification process. Some of them are discussed in the upcoming sections. 
     
\subsection{Limitations of Common Criteria}
In addition to the many advantages of CC, there are also several limitations, including
\begin{figure}[]
        \centering
        \includegraphics[scale=0.048]{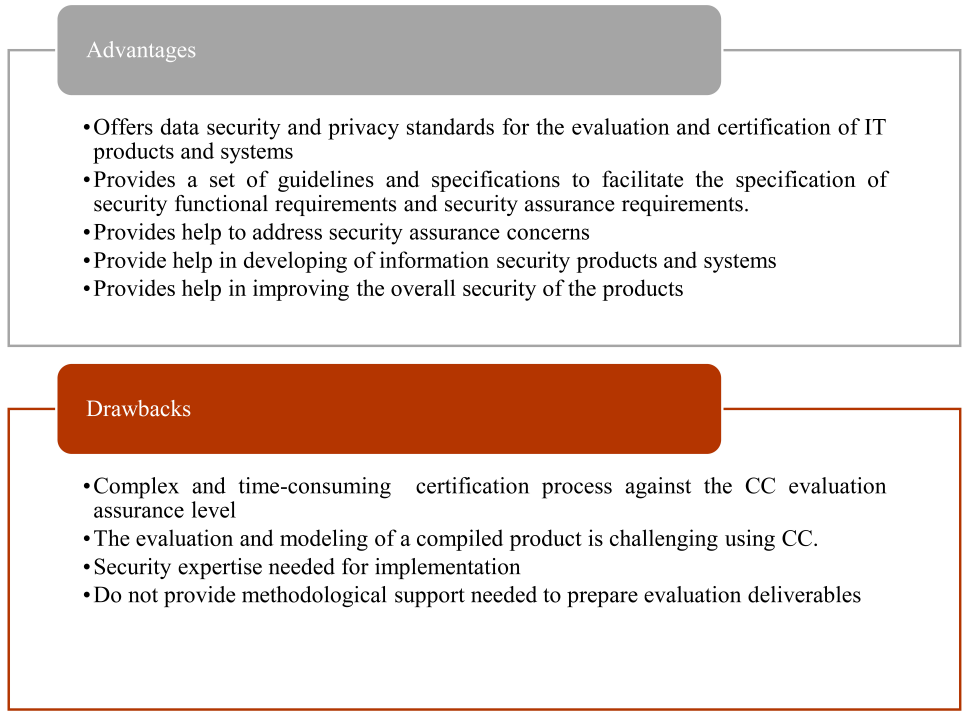}
        \caption{CC advantages and drawbacks.}
        \label{fig:my_label}
   \end{figure}
\begin{enumerate}[\itshape(a)]
    \item \textit{Complex Certification Process}: The certification process of a product or a system against the CC evaluation assurance level is very complex and time taking particularly for higher evaluation assurance levels \cite{ekclhart2007ontological}.
    
 \item \textit{Evaluation and Modelling of Composed Systems}: The CC  has been focused on a specific product that is made up of a single software component. It is needed to provide methodological support for the evaluation and modelling of security of composed products composed of two or more evaluated or unevaluated components \cite{kou2008modeling}. 
    
 \item \textit{Security Expertise Needed for Implementation}: The security standard ISO 15408 CC provides support in developing a secure system in terms of the knowledge, security expertise, and guidelines needed including secure design technique such as UMLsec. However, CC has formulated the security  guidelines and expertise in security domain terminology, making it difficult for non-security developers or stakeholders to understand. Therefore, some general security and design experience is required to get complete benefit of the CC \cite{houmb2010eliciting}.

 \item	\textit{Lack of Methodological Support for Preparing Evaluation Deliverable}: For CC-based IT products and systems security evaluation requires  evaluation deliverable such as development documents that consist of functional specification and high-level design, and operational documents that consist of guidelines documents for users and administrators, and vulnerability analysis.   Preparation of evaluation deliverable at the later stage of the development or after the product is developed may require extra costs and time.  In evaluation deliverable preparation, CC does not provide any support in terms of methodology \cite{kim2004case}.     
\end{enumerate}

A summary of advantages and drawbacks related to the CC evaluation framework is given in Fig. 5.

   \section{Security Assurance: Challenges and Gaps}
As discussed in this section, the existing literature addresses the following challenges and gaps: 

\subsection{	Elucidation, Modelling, and Validation of Security Requirements}    
The specified security requirements may often be violated after the implementation phase because of improper deployment, change in requirements, hazardous environment, or invalidation of the assumption under which the security requirements were specified. Therefore, an approach is needed to complement security requirements engineering methodologies to check if the security requirements elucidated in the development phase are implemented correctly \cite{ouedraogo2010agent}.  Several methods focused on elicitation and modelling the security requirements in the early development phase of the system. However, these methods have not been widely adopted because they are not easy to apply to the industries due to mismatch between the current development process and these methodologies. These methodologies are also very complex and do not provide documentation process of security properties of the IT systems \cite{taguchi2010aligning}. 
    
\subsection{Security Assurance of Composed System}
Security assurance measurement of a complex software system is essential but not always possible. The modern software systems are composed of several components such as servers and clients, protocols, and services. Security weakness or vulnerability in any of these components may compromise the entire system. Therefore, a process or methodology is required to assure the security of software components used in a wide range of applications \cite{kim2004framework, kim2005security}.  However, considering the relationship between these entities, a reverse process is also needed to combine the security values of the decomposed entities to obtain the security of the entire system \cite{pham2007security}. 

\subsection{Security Assurance in Operational Phase} The existing offline security assurance evaluation approaches to measure and evaluate cyber security are not effective and widely accepted approaches as it does not provide continuous security assurance assessments for complex operational software systems. Therefore, a process, method, or tool is needed for the operational security assurance assessment \cite{haddad2011operational,ouedraogo2012appraisal}.

\subsection{Service-Oriented Security Assurance} Service-oriented architectures decomposes the process into services hosted on the platform, which can adapt to changing load and performance requirements, and allow users to dynamically discover and use sub-services. However, security is not considered in the selection of services and sub-services. Therefore, service-oriented assurance is required to assure the security assurances of services as well as assess the security of sub-services \cite{karjoth2005service}.
  
\subsection{Security Assurance Tool} Security assurance tools help to improve the system security by building security into software systems or determining how secure it is. There is a need for a security assurance tool that measures the system's security level so that it can be improved and maintained overall \cite{pham2008near, deveci2015model}. 
\subsection{ CC Protection Profile for Trusted Computing Features} A protection profile is a document that assists in formulating a set of objectives and requirements for a specific category of products based on CC. Protection profile products can be validated and certified against the protection target. There exist some protection profiles for secure operating systems; however, no appropriate protection profile is available that considers trusted computing features such as trusted channels, trusted boots, and sealing \cite{lohr2009modeling}.   

\subsection{Automation of Security Assurance} Automated information security analysis, validation, evaluation, and testing approaches are required to obtain the evidence regarding security strength or security performance in the software products and telecommunication system \cite{savola2009software}. On the other hand, automation of the security assurance process in open source software is also essential. Since open-source software is subjected to frequent updates, therefore automation process should be able to incorporate these updates \cite{rauf2017towards}.

\subsection{Identification and Prediction of Security Vulnerability} Early identification of security vulnerabilities in the source code is an essential and challenging task in the software development process. Vulnerabilities can affect the system's confidentiality, integrity, and availability and thus cause severe damage to an organization \cite{hovsepyan2016newer}.


\subsection{Security in System Development Life Cycle}
Security in the entire software development process, starting from the requirement engineering to its final deployment, needs to be considered \cite{khan2018preliminary}.  On the other hand, the security assurance throughout the system development life cycle is also important \cite{vivas2011methodology}.
\begin{enumerate}[\itshape(a)]
    \item \textit{Security Design and Verification:}   
During the development life cycle, poor design practices such as the improper design of security functionality is a big security concern. Therefore, a process or a tool is required to design and develop a secure software system.  On the other hand, verification and certification of designs and codes are also crucial \cite{deveci2015model}.
\item	\textit{Security Assurance of Access Control Enforcement Code: }  
Security assurance is an essential property of the application code that has not been addressed before. It is important to ensure that the code behaves with the access control policy consistently \cite{pavlich2010framework}.
\end{enumerate}

\subsection{Cloud Security Assurance}
Cloud computing is the most enticing technology which offers economical and technological benefits in the different service provisioning domains. However, the increasing popularity of cloud services comes with concerns about the security assurance of its different services. Enforcement of security properties in a cloud is a challenging task. There are different security-related challenges that cloud service providers (CSPs) or cloud service customers (CSCs) face, such as

\begin{enumerate}[(a)]
    \item  \textit{Security Assurance Evaluation}
    
Businesses or organizations want to be assured that the cloud platform on which their infrastructure will be deployed is secure and will remain secure. Moreover, CSCs need to trust the CSPs with confidentiality, integrity, availability, and auditing in the cloud. Therefore, a security assurance methodology is required in order to obtain firm evidence that the security requirements of the companies are well defined and enforced \cite{bousquet2015enforcing}. There is a need for a method for both CSCs and CSPs to evaluate and compare the security assurance of offered services either qualitatively or quantitatively. It will enable CSCs to choose appropriate cloud services and CSPs to improve their service to gain better trust and meet customer security needs \cite{modic2016novel}.

\item \textit{Security Assurance of Multi-cloud applications} 

Security assurance of the multi-cloud applications, which consume and orchestrate services from multiple independent CSPs, is a challenging and unsolved issue \cite{rios2017dynamic}. 

\item \textit{Security Controls: Implementation and Effectiveness}

 Cloud ecosystems employ a variety of security controls to ensure security and privacy. Security properties that have been enforced in the cloud environment must be effective. However, it is a challenging task to measure their effectiveness in operation. Therefore, a method is required which can assure whether security controls are adequate and appropriate for specific cloud ecosystems \cite{formoso2015evidence}. 
 \item \textit{Security Monitoring and Analysis}
 
 A monitoring methodology is  required to monitor the security \cite{bobelin2015autonomic}. Moreover, an analysis tool is required for systematic security analysis of the critical cloud services \cite{ardagna2018case}. 
 \item \textit{Security Transparency and Auditing
}

Security transparency and auditing are two other essential factors that industries must consider to maintain and increase trust in offering services. Lack of security transparency in cloud-based services demotivates organizations from embracing the technology. The existing methods do not provide a definite method that helps in achieving security transparency as per users’ requirements \cite{ismail2020unified}. 

A summary of challenges and gaps in system security assurance in given in Table 2. 
\end{enumerate}
\begin{table}
\caption{Challenges and gaps in system security assurance}
\scriptsize
\centering
\begin{tabular}{|p{.03\textwidth}|p{.15\textwidth}|p{.24\textwidth}|p{.30\textwidth}|p{.14\textwidth}|p{.04\textwidth}|} 
\hline
S.N.                        & Category                                  & Challenges and Gaps                                                            & Descriptions                                                                                                                                                                                                                                                       & SDLC                                & Paper                                                                                                                                 \\ 
\hline
\multirow{5}{*}{1.~~~~~~ ~} & \multirow{5}{*}{Security assurance}       & Security assurance of
  the composed systems                                   & -A process or methodology is required to assure the security of software components  - A reverse process is needed to combine the security values of the decomposed entities to obtain the security of the entire system & Operational phase                   & \cite{kim2004framework,
  kim2005security} \cite{pham2007security}.                                 \\ 
\cline{3-6}
                            &                                           & Security assurance in
  the operational phase ~                                & Offline security
  assurance evaluation approaches are not effective and do not provide continuous
  security assessments ~                                                                                                                                        & Operational phase                   & \cite{haddad2011operational,ouedraogo2012appraisal}.                                                                 \\ 
\cline{3-6}
                            &                                           & Service-oriented security
  assurance                                          & Existing methods do not
  consider security in the selection of services and sub-services.                                                                                                                                                                         & Operational phase                   & \cite{karjoth2005service}.                                                                                           \\ 
\cline{3-6}
                            &                                           & Security assurance tool                                                        & Need for a tool for security
  assurance                                                                                                                                                                                                                           & Operational phase                   & \cite{pham2008near,
  deveci2015model}.                                                                              \\ 
\cline{3-6}
                            &                                           & Automation of security assurance                                               & Need of automated
  information security analysis, validation, evaluation, and testing approaches                                                                                                                                                                  & Operational phase                   & \cite{rauf2017towards}                                                                                               \\ 
\hline
\multirow{3}{*}{2.~~~~~~ ~} & \multirow{3}{*}{\makecell{Security \\ Requirements}}    & Elucidation, modelling,
  and validation of security requirements              & -To verify that the security requirements elucidated in the development phase are implemented correctly -Existing methods are not widely accepted and difficult to apply                                               & Development and
  operational phase & \begin{tabular}[c]{@{}l@{}}\cite{ouedraogo2010agent} \\\cite{taguchi2010aligning}. ~~\end{tabular}  \\ 
\cline{3-6}
                            &                                           & Design of security
  functionality                                             & -Improper design of
  security functionality                                                                                                                                                                                                                       & Design phase                        & \cite{deveci2015model}.                                                                                              \\ 
\cline{3-6}
                            &                                           & Security assurance of
  access control enforcement code                        & -To ensure that the
  code behaves with the access control policy consistently                                                                                                                                                                                     & Coding phase                        & \cite{pavlich2010framework}.                                                                                         \\ 
\hline
3.~~~~~~ ~                  & Protection Profile                        & CC protection profile
  for trusted computing features                         & No appropriate
  protection profile is available that considers trusted computing features
  such as trusted channels, trusted boots, and sealing                                                                                                                  & Operational phase                   & \cite{lohr2009modeling}.~~                                                                                           \\ 
\hline
4.~~~~~~ ~                  & Vulnerability Analysis                    & Identification and prediction
  of security vulnerability                      & -Early identification
  of security vulnerabilities in the source code                                                                                                                                                                                             & Design phase                        & \cite{hovsepyan2016newer}                                                                                            \\ 
\hline
\multirow{6}{*}{5.~~~~~~ ~} & Cloud Security Assurance & \multirow{2}{*}{Security assurance evaluation}                                 & A methodology is
  required to obtain firm evidence that the security requirements are well
  defined and enforced                                                                                                                                                 & Operational phase                   & \cite{bousquet2015enforcing}.                                                                                        \\ 
\cline{4-6}
                            &                                           &                                                                                & Need for a method for
  both CSCs and CSPs to evaluate and compare the security assurance of offered
  services either qualitatively or quantitatively                                                                                                             & Operational phase                   & \cite{modic2016novel}                                                                                                \\ 
\cline{3-6}
                            &                                           & Security assurance of
  multi-cloud applications                               & The security assurance
  of the multi-cloud applications is a challenging and unsolved issue. ~                                                                                                                                                                    & Operational phase                   & \cite{rios2017dynamic}.                                                                                              \\ 
\cline{3-6}
                            &                                           & Implementation and effectiveness
  of security control                         & -To measure effectiveness of security control in operation is challenging.  -A method is required which can assure whether security controls are adequate and appropriate                                                & Operational phase                   & \cite{formoso2015evidence}                                                                                           \\ 
\cline{3-6}
                            &                                           & Security monitoring and
  analysis                                             & A monitoring methodology/tool
  is required to monitor the security                                                                                                                                                                                                &Operational   phase                                 & \cite{bobelin2015autonomic}                                                                                          \\ 
\cline{3-6}
                            &                                           & Security transparency and auditing  & -Lack of security transparency and auditing in cloud-based services  -Existing methods do not support in achieving security transparency as per users’ requirements                                                   & Operational   phase                      & \cite{ismail2020unified}                                                                                             \\
\hline
\end{tabular}
\end{table}
    \section{Security Goals, Requirements and Metrics}

The security assurance evaluation process requires specific assurance goals, security requirements, and design guidelines, which can be used by security personnel to assess and ensure a high level of security assurance.  In this section, we discuss the security goals, requirements and security metrics. 

\subsection{Security Goals}    
     Organizations implement a security policy to impose a uniform set of rules to handle and protect the crucial information of the system. Most of these security policies consider three significant aspects of their data and information: confidentiality, integrity, and availability. However, these security requirements can be emphasized based on application domains. Most of the work primarily focused on these three security requirements in literature. However, some of the works also included other security goals for security assurance, such as privacy, authenticity, accountability, conformity, utility, possession, non-repudiation, and authorization, as shown in Table 3.     
  \subsection{Security  Requirements }

Adequate security assurance signifies that a software system's specified and predefined security assurance requirements have been satisfied during the security assurance assessment processes and activities. Security requirements can be categorized as functional and non-functional requirements that need to be satisfied to achieve the security attributes of a software system. The functional requirements define what the system does or must not do; it must be testable, which means the requirement can be tested to check whether it is fulfilled. Non-functional requirements define how the system should do. It considers the performance of the whole system. Security assurance requirements can be determined by analyzing the security requirement of a software system, security policies, business drivers, operational environment, etc. The security requirements should be based on the iterative threat, vulnerability, and risk analysis and should also incorporate the technical and architectural information \cite{savola2009software}.  

\begin{table}[]
\scriptsize
\centering
 \caption{Summary of security goals considered in the literature.}
\label{tab:comparison}
\begin{tabular}{|p{.15\textwidth}|p{.60\textwidth}|}
\hline
Security Goals  & Papers \\ \hline
Confidentiality &    \cite{tashi2010security, vivas2011methodology,bousquet2015enforcing, deveci2015model,ardagna2018case,hudic2017security}

\cite{jeong2005framework,naeem2019framework}  \cite{al2009non}  \cite{savola2009software}  \cite{savola2010towards}   \cite{sklyar2017challenges} \cite{ismail2020unified}   \cite{katt2018quantitative} \cite{agrawal2019multi}
    \\ \hline
Integrity       &   \cite{tashi2010security, vivas2011methodology,bousquet2015enforcing, deveci2015model,ardagna2018case,hudic2017security}
\cite{jeong2005framework,naeem2019framework}  \cite{al2009non}  \cite{savola2009software}  \cite{savola2010towards}   \cite{sklyar2017challenges} \cite{ismail2020unified}   \cite{katt2018quantitative} \cite{agrawal2019multi} \cite{gutierrez2015business} 
     \\ \hline
Availability    &    \cite{tashi2010security, vivas2011methodology,bousquet2015enforcing, deveci2015model,ardagna2018case,hudic2017security}
\cite{jeong2005framework,naeem2019framework}  \cite{al2009non}  \cite{savola2009software} \cite{savola2010towards}   \cite{sklyar2017challenges} \cite{ismail2020unified}   \cite{katt2018quantitative} \cite{agrawal2019multi} \cite{ouedraogo2008deployment,ouedraogo2010agent}  \cite{gutierrez2015business} 
    \\ \hline
Privacy       & \cite{gutierrez2015business}        \\ \hline
Authenticity    &  \cite{jeong2005framework,naeem2019framework} \cite{al2009non}  \cite{savola2009software}  \cite{savola2010towards}   \cite{sklyar2017challenges} \cite{gutierrez2015business}  \cite{rauf2017towards}      \\ \hline
Accountability  &    \cite{ismail2020unified}   \cite{katt2018quantitative} \cite{agrawal2019multi}    \\ \hline
Conformity      & \cite{ismail2020unified}   \cite{ouedraogo2008deployment,ouedraogo2010agent}        \\ \hline
Utility         &     \cite{al2009non}    \\ \hline
Possession      &  \cite{al2009non}       \\ \hline
Non-Repudiation & \cite{savola2009software}  \cite{sklyar2017challenges} \cite{agrawal2019multi}       \\ \hline
Authorization   &     \cite{savola2009software}  \cite{savola2010towards}   \cite{rauf2017towards}   \\ \hline
\end{tabular}
\end{table}


CC provides a structured method to list the security requirements that include an IT system's functional and assurance requirements. Security function requirements describe various functional requirements in terms of communication security, security audits, data protection, authentication, security management, system access, trust path, etc. System developers can select a subset of these requirements to implement in the form of security properties as a part of the security target document of the TOE. On the other hand, security assurance requirements are the requirements that need be fulfilled to assure that the security functions are implemented correctly. Security assurance requirements cover configuration management, guideline documents, delivery and operation, assurance test, vulnerability assessment, etc \cite{debbabi2007embedded, knapp2014industrial, chapple2018isc}. 

  Various methods have been developed in the literature for security requirements elicitation, tracking, analysis, correctness, modelling, etc., as shown in Fig. 6. Some of these methods are:
  
\begin{figure}
        \centering
        \includegraphics[scale=0.09]{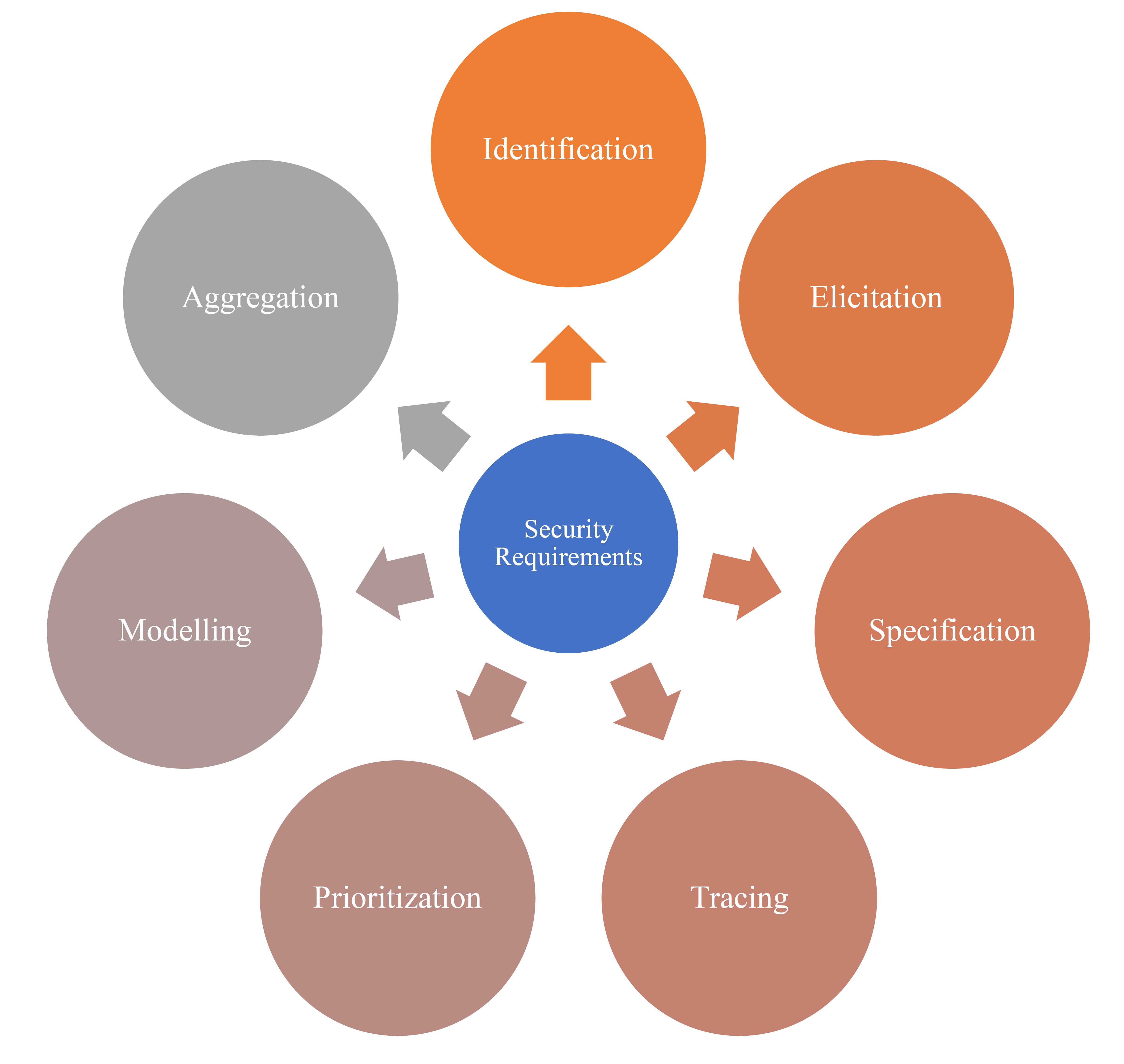}
        \caption{Security requirements methods.}
        \label{fig:my_label}
 \end{figure}

\subsubsection{Elicitation, Tracing, and Analysis of the Security Requirements}
      Building a secure system is complicated for several reasons, including a lack of security expertise in development teams, inadequate methodologies to support non-expert developers, etc. The security standard such as CC can be used to specify the security functional and assurance requirements, and UMLsec can be used for model-based security engineering. However, it is difficult to understand this security expertise and guidelines for the developers because it is not written explicitly. Therefore, a methodology is required for elicitation, tracking, and analyzing the security requirement. Houmb et al. \cite{houmb2010eliciting} developed such a methodology called SecReq by combining three techniques: CC, the heuristic requirements editor HeRA, and UMLsec. The SecReq is designed to make the security requirements engineering more systematic and effective by integrating elicitation, traceability, and analysis activities.

\subsubsection{ Security Requirements Modelling} 

     Eliciting and modelling the security requirements in the development phase are well-known practices to prevent potential vulnerabilities. However, the existing methodologies such as KAOS \cite{darimont1997grail}, SecureTropos \cite{mouratidis2007secure} are neither useful nor widely accepted in the industries because of mismatching in the software development process and their complexity. There is also a lack of a standard for documentation of the security properties of a software system concisely and systematically during the development process.   
 Taguchi et al. \cite{taguchi2010aligning}  developed a framework that provides a security requirement modelling method for the system development and security assurance under the CC. This framework aligned the security requirements and assurance in a single requirement modelling methodology uniformly and concisely.

\subsubsection{Correctness of the Security Requirements}    

In general, the security requirements are identified during the design phase based on the assumption made on the system operating environment. These assumptions are no longer valid if there are any changes in the system environment. On the other hand, security requirements may be adequately identified during the development phase but not correctly deployed or become less effective due to unidentified hazards in the system.   Due to this, it is not easy to ensure that the secure system will remain secure over time. Therefore, an approach is needed for continuous security assurance, which also supports whether the security requirements can be fulfilled during the system operation based on the evidence collection.    Ouedraogo et al. \cite{ouedraogo2010agent} developed an approach that complements security requirements engineering methodologies to check whether the security requirements elucidated during the development phase of the system have been correctly implemented by collecting continuous evidence. 

\subsubsection{Measurement Requirements }
   
The continuous independent evolution of the complex and operating system components makes the security measurements more challenging. In the context of security assurance, the correctness of the security controls is the main objective of the measurements. Security measurements and their different properties change over time. Therefore, the measurement framework should consider the variation in the measurement target and available measurement infrastructure. With the evolvement of the available measures, it is vital to manage the dynamic features. Kanstrén et al.   \cite{kanstren2010towards} introduced a taxonomy-based approach and proposed an  abstraction Layer between the measurements identified in the measurements framework based on the measurement infrastructure and requirements. This approach helps to relate the available and achievable measurements to the measurement requirements of security assurance plans and in managing the dynamic features in measurement requirements. 

     \subsection{Security Metrics }
The dynamic nature and complexity of the security risk make it challenging to measure security as a universal property. Lack of standard definition is also one of the main reasons behind this. Metrics is the widely used and more suitable term for security-related objectives \cite{savola2010towards}. Security metrics offer security-related information from a different point of view which helps in essential and credible information security measurement of a software system. In the literature, extensive research works have been done on defining the metrics taxonomy. NIST \cite{swanson2003security} provides standards for determining the adequacy of in-place security methods, policies, and procedures using metrics. It explains how metrics can be developed and implemented and how they may be used to justify the security procedures investment.
\subsubsection{Quality Criteria of Security Metrics}

 The three core quality criteria of security metrics are correctness, measurability, and meaningfulness. These criteria are crucial for security metrics measurements and their practical application. To ensure the correctness of the security metrics, a well-established and systematic development methodology is required, which includes validity and reliability analysis; measurability of the security metrics can be improved by continuous development of an efficient and relevant measurement framework for the system under evaluation \cite{savola2011visualization}. Moreover, the simplicity and reliability of the metrics are also essential to make the security assurance methodology more robust. It enables a fast and objective evaluation of a system's security assurance level. \cite{ouedraogo2009risk}.  Several works have been done in the literature on methods and quality parameters of security metrics, as shown in Table 3, Fig. 7 and Fig. 8. 

\begin{figure}
        \centering
        \includegraphics[scale=0.11]{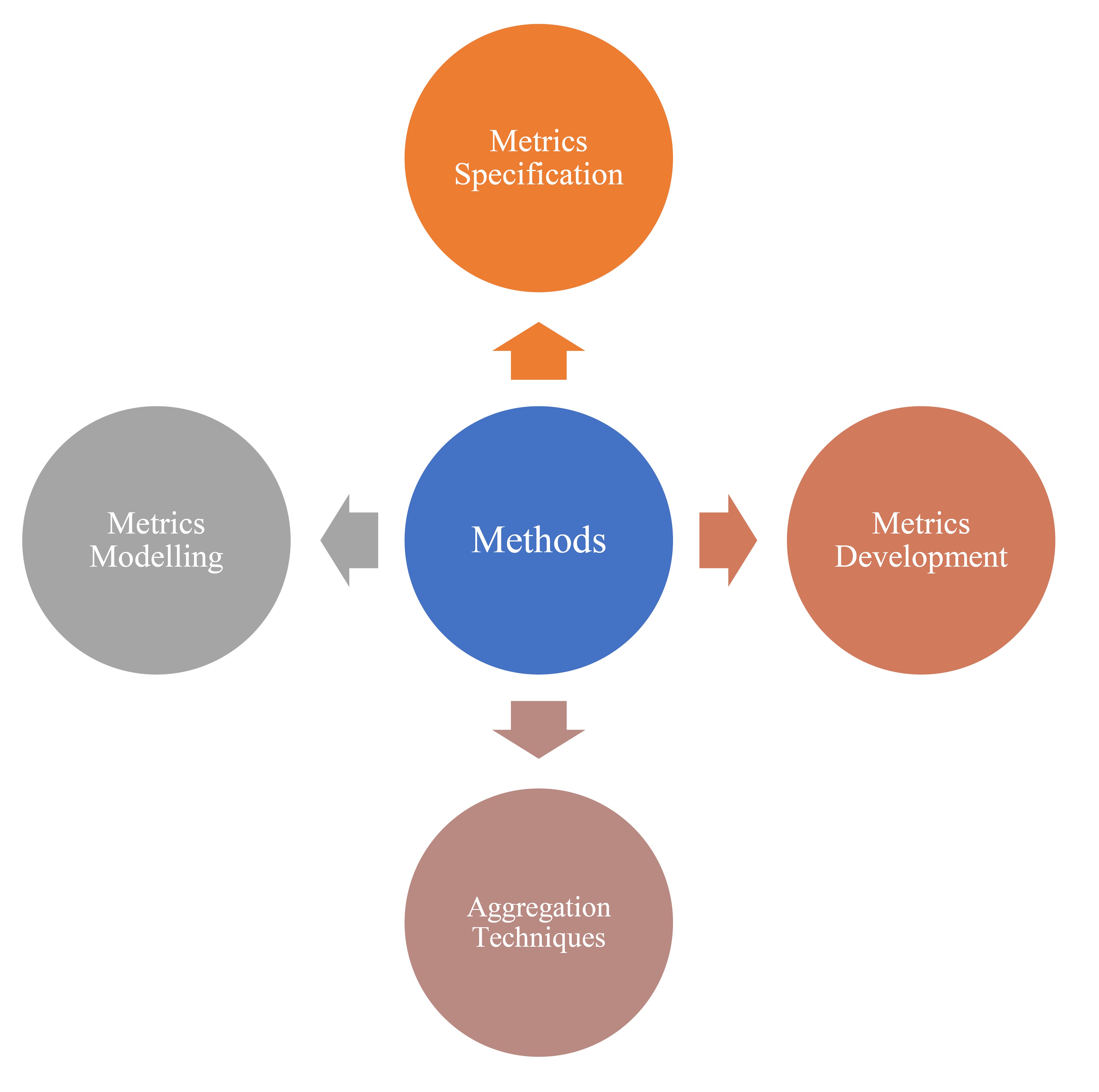}
        \caption{Security metrics methods.}
        \label{fig:my_label}
 \end{figure}
\begin{figure}
        \centering
        \includegraphics[scale=0.14]{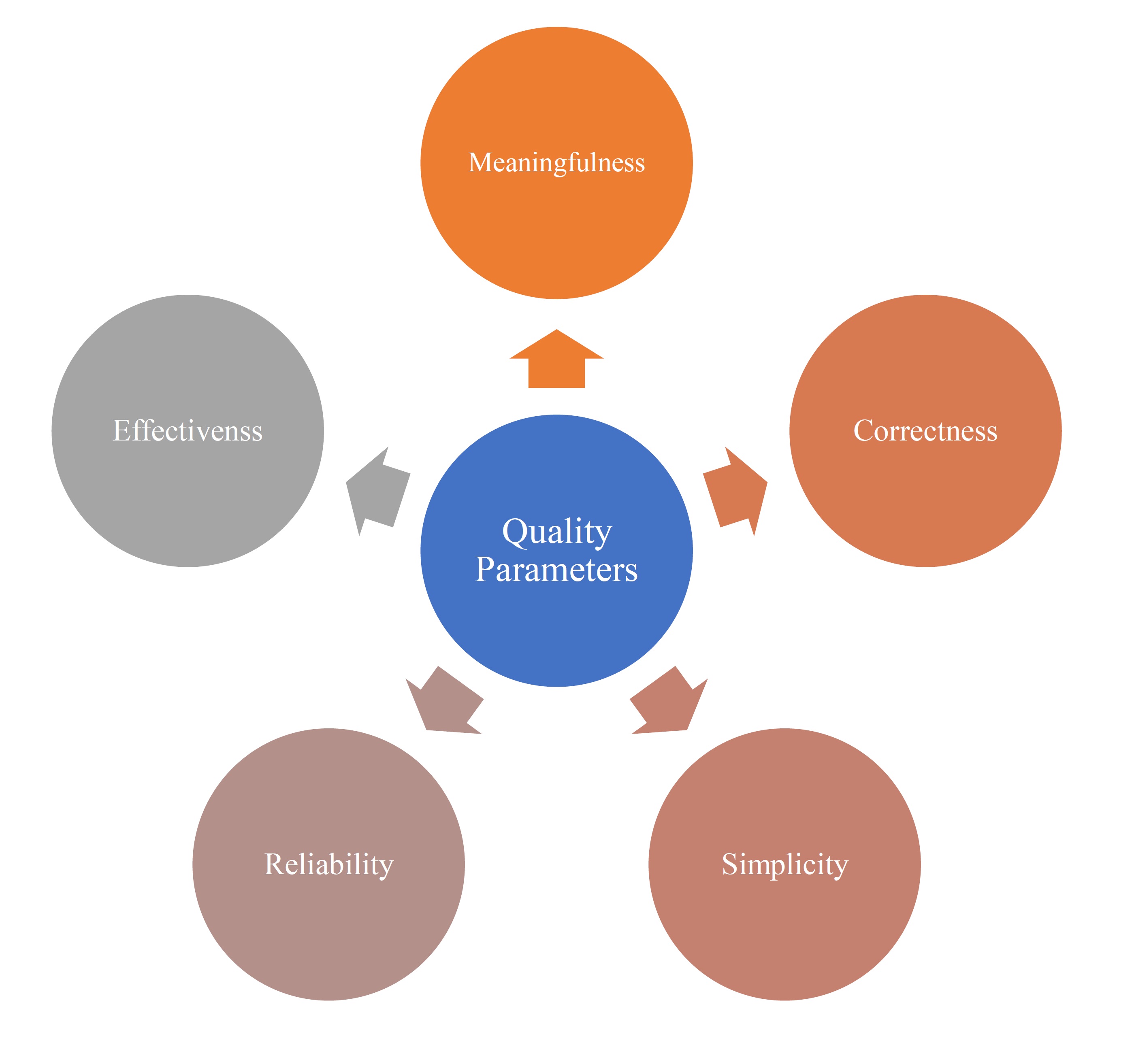}
        \caption{Security metrics methods.}
        \label{fig:my_label}
 \end{figure}

\begin{table}[]
\scriptsize
\centering
\caption{Summary of security metrics category, goals, and methods.}
\begin{tabular}{|p{.03\textwidth}|p{.12\textwidth}|p{.18\textwidth}|p{.18\textwidth}|p{.10\textwidth}|p{.20\textwidth}|p{.04\textwidth}|}
\hline
S.N.               & Category                                                       & Goals                                                                                                            &\pbox{20cm}{ Method/ \\Technique}                                                                                                                  & Domain                               & Description                                                                                                                                                                                                                                                    & Paper                                                                                                                                                  \\ \hline
1.                  & Security Metrics   Development                                 & Development of security   effectiveness metrics for ensuring the correctness of security controls.               & Risk-Driven Security   Metrics Development Methodology                                                                             & IT Systems (a Push   E-mail service) & In this approach,   security control effectiveness is measured as part of the assurance of   security control correctness at the higher level.                                                                                                                 &                  \cite{savola2010towards}                          \\ \hline
2. & \multirow{4}{*}{\pbox{20cm}{Security \\ Metrics  \\ Measurement, \\ and \\Management}} & Security correctness,   security effectiveness and the quality of the security verification process   at runtime & Developed security   assurance monitoring tool and the measurement framework                                                       & \pbox{20cm}{IT \\Systems}                           & Authors proposed a set of   metrics for the appraisal of security assurance at runtime.                                                                                                                                                                        &       \cite{ouedraogo2011new}                                                                                             \\ \cline{3-7} 
                   &                                                                & Verification of the   security mechanisms at runtime                                                             & Developed a taxonomy of   quality metrics                                                                                          & \pbox{20cm}{IT \\Systems}                             & Authors discussed the   impact of various properties on the confidence of the measurement data, such   as trusted platform module for measurement data and infrastructure assurance,   trusted monitoring base techniques, and measurement probe form factors. &        \cite{ouedraogo2013taxonomy}                  \\ \cline{3-7} 
                   &                                                                & Specification and measurement   of security metrics                                                              & Developed visualization   and modeling tool                                                                                        & Software Systems                     & This tool helps to   increase the meaningfulness of metrics in security assurance and risk   management contexts by hierarchical metrics modelling.                                                                                                            &                                \cite{savola2011visualization}                                    \\ \cline{3-7} 
                   &                                                                & Evaluation of IT   systems security assurance.                                                                   & Security assurance   metric and aggregation techniques                                                                             & \pbox{20cm}{IT \\Systems}                            & Authors developed a risk-based security   assurance metric and aggregation technique  for evaluating the systems security   assurance.                                                                                                                         &                       \cite{ouedraogo2009risk}                                                  \\ \cline{3-7} 
                   &                                                                & Confidence in   measurement data during operational security assurance                                           & Describes the   properties of a trusted measurement base and implemented as a part of metrics   visualization tool and prototyped. & Cloud computing                      & Authors discussed the   impact of various properties on the confidence of the measurement data, such   as trusted platform module for measurement data and infrastructure assurance,   trusted monitoring base techniques, and measurement probe form factors. &  \cite{kanstren2015security} \\ \hline
\end{tabular}
\end{table}

    
 \subsubsection{   Security Metrics Management and Measurement  }
 This subsection discusses various factors and techniques for security metrics management and measurement.
\begin{enumerate}[(a)]
    \item \textit{ Meaningfulness of Security Metrics and Measurements} 
    
    The meaningfulness of security metrics and measurements is remarkably challenging in security decision-making, such as risk management and security assurance. Because of poor management and a significant number of uncategorized data, many security metrics activities have a low level of meaningfulness. Primarily, only a limited number of metrics have been developed because they are more understandable in decision-making than a large number of metrics. However, much essential information related to security can be lost in the aggregation process of the low-level metrics. Therefore, systematic and complete security metrics management and measurement are essential. Metrics visualization facilitates the management and measurements of the security metrics to enhance the meaningfulness of the decision-making process. A visualization and modelling tool is developed by Savola and Heinonen \cite{savola2011visualization} for hierarchical specification and deployment of the security metrics and their measurements. This tool helps to increase the meaningfulness of metrics in security assurance and risk management contexts by hierarchical metrics modelling. It also connects high-level security objectives with detailed measurements.

    \item \textit{Confidence in Security Metrics Measurement }     
    
The data collected during the system's operation helps in expressing the system's current state and validating the security metrics model.   Therefore, one should have confidence in security metrics and data measurement. However, several factors can impact confidence, and trust is one of these significant factors.   Kanstrén, T., \& Evesti \cite{kanstren2015security} discussed the impact of various properties on the confidence of the measurement data, such as trusted platform module for measurement data and infrastructure assurance, trusted monitoring base techniques, and measurement probe form factors. They also defined a set of measurement data confidence levels based on the trusted monitoring base achieved. They implemented this approach as part of a metrics visualization tool in a private cloud environment. 

\item \textit{Security Metrics Aggregation}

 An efficient and straightforward aggregation method to combine the security assurance of sub-components by considering their relationship is essential for a robust security assurance methodology. A risk-based security assurance metric and aggregation technique is proposed by Ouedraogo et al. \cite{ouedraogo2009risk} that can be combined with a methodology for evaluating the systems security assurance. 

\end{enumerate}
\subsubsection{ Security Metrics Model for Operational Security }
To measure the operational security assurance of a system, it is essential to understand and express the current and anticipated security posture of the system. Security metrics modelling is a significant way to express the security status of the system \cite{kanstren2015security}. The data collected during a system's operation is used to express the system's current state and validate the security metrics model.

 \subsubsection{ Security Metrics Taxonomy for Run-time Systems     }

Ouedraogo et al. \cite{ouedraogo2012appraisal,ouedraogo2011new, ouedraogo2013taxonomy} made an effort to develop a set of metrics to evaluate the security assurance of the runtime systems. Ouedraogo et al. \cite{ouedraogo2013taxonomy}  developed metrics taxonomy based on CC and the System Security Engineering Capability Maturity Model (SSE-CMM). They represented the verification probe quality levels based on five capability maturity levels of the  SSE-CMM and  some of the CC families such as scope, depth, rigour, and independence of verification) capabilities as requirements to attain a defined level of quality. They also made an analysis of the mapping between different capability levels and the quality levels of the different verification metrics families, such as coverage, rigour, depth, and independence of verification.   Ouedraogo et al. \cite{ouedraogo2012appraisal,ouedraogo2011new} also developed a method to combine the security metrics into the quantitative or qualitative indicators that are crucial in developing understanding regarding the security status of an IT system component.

A summary of security metrics category, goals, and methods is given in Table 4.


\section{Security Assurance Methods  }

Security assurance involves demonstrating with evidence that a system fulfills established standard security criteria \cite{kim2004case}. It provides the confidence that a system meets the security requirements and therefore has fewer vulnerabilities, resulting in the overall reduction of risk in using the related system. Due to the importance of security assurance, most organizations make the efforts to demonstrate the security assurance of their systems but mostly rely on less effective approaches \cite{kim2004framework}. To enhance security assurance of the systems,various methods have been developed, as shown in Figure 9 and outlined in the upcoming subsections. A categorical representation of security assurance methods is given in Table 5. This table has categorized the security assurance methods based on the various system development phases such as governance, construction, and deployment. The governance considers policy and compliance, strategy \& metrics, risk, and awareness. The construction phase considers requirements, design/modelling, and verification, and the deployment includes monitoring and management. On the other hand, various methods and techniques have been used in developing security assurance process or methodology       

A security assurance method can be classified based on its qualitative or quantitative nature. Qualitative methods use the ordinal rating scale ( e.g., 1-5, low, high, critical) to represent the security level of a software system. On the other hand, quantitative methods use factual and measurable data to calculate the security level of a software system. This method is mainly based on mathematical and computational techniques. Fig. 10 represents the evolution of these two methods. This figure depicts that major of the researches have been focused on the qualitative security assurance methodology, and very few efforts have been made toward developing a quantitative security assurance methodology. Katt and Prasher \cite{katt2018quantitative} discussed the advantages of the quantitative security assurance method over the qualitative security assurance method. They discussed the advantages of quantitative security assurance metrics, including both positive and negative security aspects. 
 \begin{figure}
        \centering
        \includegraphics[scale=0.075]{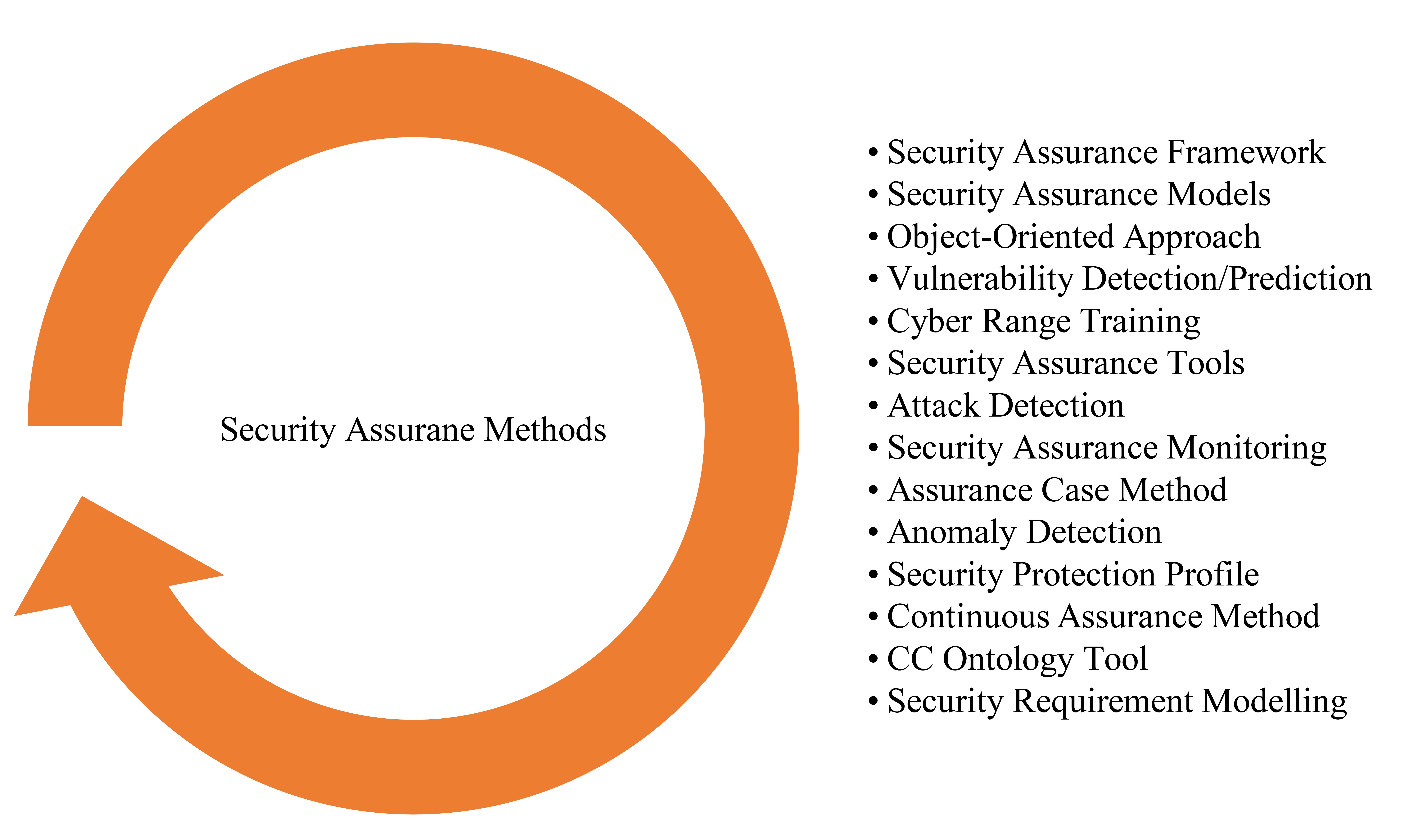}
        \caption{Security assurance methods.}
        \label{fig:my_label}
     \end{figure}
 \begin{figure}
        \centering
        \includegraphics[scale=0.56]{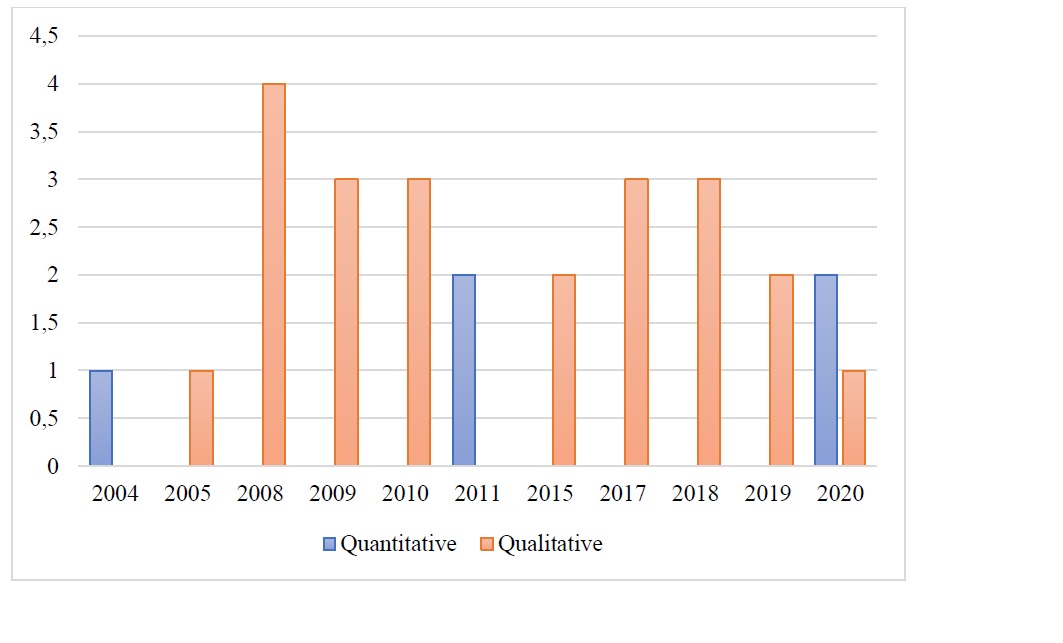}
        \caption{Quantitative and qualitative security assurance.}
        \label{fig:my_label}
     \end{figure}

In the literature, various methods and techniques have been used for security assurance as shown in Figure 11.
    
 \begin{figure}
        \centering
        \includegraphics[scale=0.24]{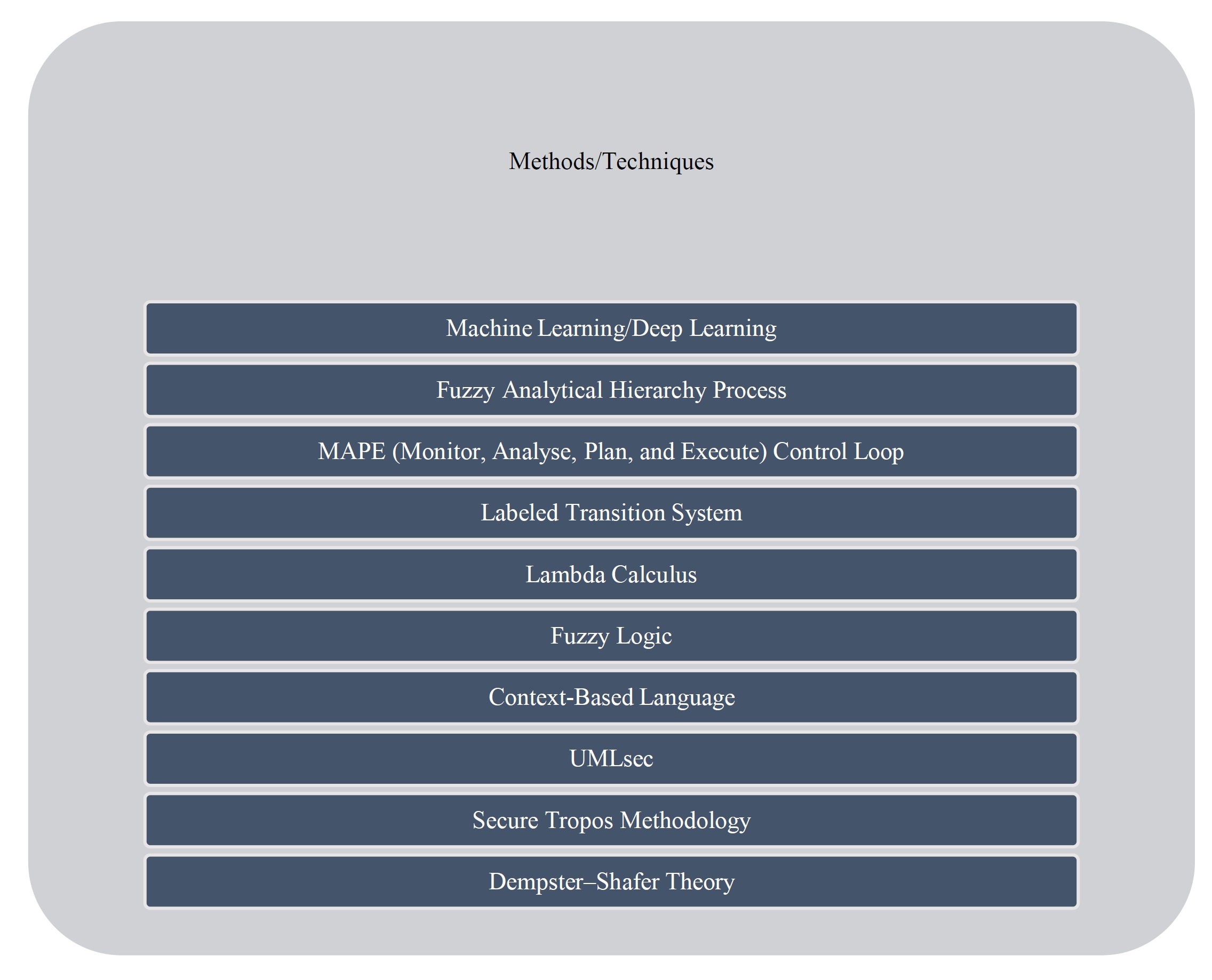}
        \caption{Methods and techniques used for security assurance.}
        \label{fig:my_label}
     \end{figure}



\begin{table} [!h]
\scriptsize
\centering
\caption{Security assurance methods in system development phases.}
\begin{tabular}{|p{.02\textwidth}|p{.092\textwidth}|p{.13\textwidth}|p{.20\textwidth}|p{.45\textwidth}|}
\hline
\multirow{14}{*}{
  \shortstack{ \rotatebox{90} {System
  Development}}} & \multirow{4}{*}{Governance}   & Policy and  Compliance                      & \multicolumn{2}{l!{\color{black}\vrule}}{Access
  Control Policy \cite{pavlich2010framework}   }                                                                                                 \\ 
\cline{3-5}
                                                                       &                               & Strategy \& Metrics
                    & \multicolumn{2}{l!{\color{black}\vrule}}{\parbox{9cm}{  Security Preference \cite{rizvi2018security}, Assurance Profile \cite{marquet2010security, haddad2011operational}, \\
  Protection Profile \cite{lohr2009modeling, bock2012towards}, Security Metrics \cite{ouedraogo2009risk, ouedraogo2012appraisal, ouedraogo2013taxonomy, savola2010towards, haddad2011operational, ouedraogo2011new, savola2011visualization, kanstren2015security, modic2016novel, rizvi2018security}}}                                                                \\ 
\cline{3-5}
                                                                       &                               & Risk
                                  & \multicolumn{2}{l!{\color{black}\vrule}}{Risk Assessment \cite{ouedraogo2011new, tanna2005information}, Risk Management \cite{savola2011visualization}, Risk
  Model \cite{ray2015security}}                                                                          \\ 
\cline{3-5}
                                                                       &                               & Awareness                                  & \multicolumn{2}{l!{\color{black}\vrule}}{Security Awareness Program \cite{gupta2020cyber},  Cyber Range
  Training \cite{gupta2020cyber}}                                                                          \\ 
\cline{2-5}
                                                                       & \multirow{7}{*}{Construction} & Requirements                              & \multicolumn{2}{l!{\color{black}\vrule}}{\shortstack[l]{\\  Elicitation \cite{houmb2010eliciting, katt2018quantitative}, 
  Specification \cite{jaiswal2017engineering, katt2018quantitative},   Identification \cite{cadzow2004securitv, katt2018quantitative},    Aggregation \cite{pham2007security, katt2018quantitative},\\ Measurement \cite{kanstren2010towards, katt2018quantitative},
  Prioritization \cite{jaiswal2017engineering},    Modelling \cite{kim2004framework, yavagal2005common, taguchi2010aligning, katt2018quantitative}, \\  Correctness \cite{ouedraogo2010agent},   Tracing \cite{houmb2010eliciting},     Security Requirements Engineering Process  \cite{mellado2007common},  \\   Requirement Representation \cite{bobelin2015autonomic, ekclhart2007ontological}        Ontological Mapping \cite{ekclhart2007ontological}     }   }                                     \\ 
\cline{3-5}
                                                                       &                               & Design/Modelling                           & \multicolumn{2}{l!{\color{black}\vrule}}{\shortstack[l]{ \\ Security Modelling \cite{ouedraogo2010agent, tashi2010security, vivas2011methodology, deveci2015model, rauf2017towards, cadzow2004securitv, kim2004framework, baldwin2006model, karjoth2005service, khan2018preliminary, wu2008network}, \\ Vulnerability
  Prediction Modelling \cite{hovsepyan2016newer},   Security Assurance Model \cite{khan2018preliminary, xiang2010analysis}, \\ Security Assurance Development Process Model \cite{lan2015sadp}, \\ Assurance Modelling of System Component \cite{ouedraogo2010information, kim2004framework}, \\ Model-Driven Security Framework \cite{deveci2015model, rauf2017towards},   Network Security Assurance \cite{bialas2019anomaly}, \\ Security Assurance Framework \cite{katt2018quantitative, diamantopoulou2020aligning},  CC-Based Model for Network Security \cite{wu2008network}, \\ Framework for Authenticity \cite{naeem2019framework},   Object-Oriented Security Assurance Model \cite{cuihua2009object}, \\ Component Security Assurnace \cite{kim2004framework, wu2008network, kim2005security, karjoth2005service},  Usable-Security Assurance \cite{agrawal2019multi}, \\ Service-Oriented Security Assurance \cite{karjoth2005service},  Formal Method    \cite{kim2003reliability, maidl2008formal}, \\  Security Assurance in Open Infrastructure \cite{ouedraogo2008deployment}, \\ Other Security Assurance Methods \cite{pavlich2010framework, sakthivel2020core, ardagna2018case, sklyar2017challenges, modic2016novel, rizvi2018security, hubballi2019cloud, hudic2017security, drago2012inside, ismail2020unified} }}  \\ 
\cline{3-5}
                                                                       &                               & \multirow{5}{*}{Verification}              & Code Review         & Vulnerability Detection Method \cite{akram2021sqvdt, dissanayaka2020security, bilgin2020vulnerability},   Vulnerability  Prediction   \cite{hovsepyan2016newer}                                                                                                                     \\ 
\cline{4-5}
                                                                       &                               &                                            & Security Testing \cite{savola2009software}    & Penetration Testing \cite{cheah2018building, katt2018quantitative}, Vulnerability
  Testing \cite{katt2018quantitative, naeem2019framework, huang2005testing, tanna2005information},  Black Box Testing \cite{krishnan2011applying, huang2005testing},  Unit Testing \cite{krishnan2011applying}, Code Inspection \cite{krishnan2011applying}, Security Requirement Testing \cite{jaiswal2017engineering, katt2018quantitative}                                           \\ 
\cline{4-5}
                                                                       &                               &                                            & Threat Assessment   & Threat Analysis \cite{tanna2005information},   Threat Modelling     \cite{cheah2018building, katt2018quantitative, cadzow2004securitv, baldwin2006model}                                                                                                                           \\ 
\cline{4-5}
                                                                       &                               &                                            & Security Assessment & System Modelling, Vulnerability Analysis \cite{savola2010towards, savola2009software} Interview \cite{knowles2015assurance}                                                                                                                    \\ 
\cline{4-5}
                                                                       &                               &                                            & \multicolumn{2}{l!{\color{black}\vrule}}{Security Review \cite{lan2015sadp}}                                                           \\ 
\cline{4-5}
                                                                       &                               &                                            & \multicolumn{2}{l!{\color{black}\vrule}}{Quality of Security verification \cite{ouedraogo2011new} }                                                                                           \\ 
\cline{2-5}
                                                                       & \multirow{3}{*}{Deployment}   & \multirow{3}{*}{\shortstack{Monitoring and \\ Management}} & \multicolumn{2}{l!{\color{black}\vrule}}{Monitoring and Auditing \cite{karjoth2005service, ouedraogo2008deployment, jahan2020mape, trapero2017novel}, Anomaly
  Detection  \cite{bialas2019anomaly}}                                                                                \\ 
\cline{4-5}
                                                                       &                               &                                            & \multicolumn{2}{l!{\color{black}\vrule}}{Assurance Management, Security
  Durability \cite{kumar2020knowledge}}                                                                                 \\ 
\cline{4-5}
                                                                       &                               &                                            & \multicolumn{2}{l!{\color{black}\vrule}}{Vulnerability Management \cite{akram2021sqvdt, dissanayaka2020security, bilgin2020vulnerability, katt2018quantitative, naeem2019framework, huang2005testing, tanna2005information, hovsepyan2016newer}}                                                                                                    \\
\hline
\end{tabular}
\end{table}


\subsection{Security Assurance Framework}
In the literature, the following security assurance frameworks have been proposed and discussed: 

\subsubsection*{ (a) Model-Driven Framework for Security Functionality Verification  }
   The major security problem for a software system is poor design practices, such as improper security functionality design and implementation in the development cycle in an improper manner. Therefore, it is crucial to verify the security functionality that is designed and developed. The codes and designs should be verified or certified by a competent authority. One of the most well-known frameworks for evaluating the security functionality of software systems is the CC framework. Deveci et al. \cite{deveci2015model} designed a  framework to assist designers, testers, and analysts during the CC certification process. The proposed framework is a model-driven security framework used to analyze, design, and evaluate the security properties of information systems. This framework also supports developers and evaluation authorities in implementing the security assurance process through formal methods based on UML, object constraint language, Promela, and Spin.

\begin{table}
\scriptsize
\centering
\caption{A summary of security assurance frameworks}
\begin{tabular}{|p{.03\textwidth}|p{.16\textwidth}|p{.22\textwidth}|p{.22\textwidth}|p{.12\textwidth}|p{.04\textwidth}|} 
\cline{1-4}\arrayrulecolor{black}\cline{5-5}\arrayrulecolor{black}\cline{6-6}S.N.      & Challenges                                                     & Solutions                                                                                                               & Framework                                                                                                   & Domain                                 & Paper                                                                            \\ 
\cline{1-4}\arrayrulecolor{black}\cline{5-5}\arrayrulecolor{black}\cline{6-6}
1. & Improper security
  functionality design and implementation    & Verification or
  certification of codes and designs                                                                   & Model-driven security
  framework to analyze, design, and evaluate the security properties of
  information systems. & Software Systems                         & \cite{deveci2015model}                                         \\ 
\cline{1-4}\arrayrulecolor{black}\cline{5-5}\arrayrulecolor{black}\cline{6-6}
2. & Inaccuracy in
  quantitative security assurance measurement    & Development of security
  assurance framework considering various security requirements, threats, and
  vulnerabilities & A quantitative security
  assurance framework incorporating both security assurance and vulnerabilities.             & IT and CPS                               & \cite{katt2018quantitative}                                     \\ 
\cline{1-4}\arrayrulecolor{black}\cline{5-5}\arrayrulecolor{black}\cline{6-6}
3. & Authenticity of the
  software application before installation & Check and ensure the
  authenticity of the software application before installation.                                    & Framework for checking
  the authenticity or credibility of applications or software                                 & Applications or
  software               & \cite{naeem2019framework}                                      \\ 
\cline{1-4}\arrayrulecolor{black}\cline{5-5}\arrayrulecolor{black}\cline{6-6}
4. & Security assurance of composed
  information security products & Security assurance
  method by analysis of potential components and interfaces between the
  components                 & CC-based model to assess the complete
  network security by combining evaluation assurance levels EAL and CAP       & Composed Information
  Security Products & \cite{kim2004framework,
  kim2005security, karjoth2005service}  \\
\cline{1-4}\arrayrulecolor{black}\cline{5-5}\arrayrulecolor{black}\cline{6-6}
\end{tabular}
\end{table}


   \subsubsection*{ (b) Quantitative Security Assurance Framework}
           
    There are many approaches to security assurance in the literature; however, measuring the security of a software system is still a complex and tricky process. These approaches do not accurately measure the system's security level because they either consider only one aspect of assurance, such as threats/vulnerabilities, or do not consider the significance of the various security requirements to the system under evaluation. Considering this fact, Katt and Prasher \cite{katt2018quantitative} developed a security assurance framework incorporating both security assurance and vulnerabilities. The proposed method is quantitative           

\subsubsection*{ (c) Security Assurance Framework for Software Authenticity}
There is a high risk of cyber-attacks on software applications because of their popularity, misconfiguration, technical flaws, and vulnerabilities. Therefore, the software that will be installed on the critical systems must be secure. Naeem et al.  \cite{naeem2019framework} considered the authenticity of the software application before its installation and proposed a framework to check the authenticity of the software application before its installation. However, tools and frameworks are available, but they consider only a single aspect, such as a specific OS or a single-entry point check. The framework proposed by Naeem et al. provides a solution to overcome this challenge.

\subsubsection*{ (d) Security Assurance Model for Composed Information Security Products}
The general systems are made with different components such as servers, protocols, services, and clients. It is important to note that any weakness in one of these components may compromise the whole system. Therefore, efficient security assurance methods are required to secure these systems. The composed assurance packages (CAPs) is an evaluation method for composed security products. However, it requires analysis of potential components and interfaces between the components, which is quite difficult because of the complexity and variation in new IT products. Wu et al. \cite{wu2008network} discussed this issue and developed a CC-based model to assess the complete network security by combining evaluation assurance levels (EAL) and CAP, where EAL is used to evaluate a single IT entity while CAP is used to evaluate the composite IT entities. Some other efforts have been made to develop component security assurance methods based on the security properties of the individual system components\cite{kim2004framework, kim2005security, karjoth2005service}. 

An overview of security assurance frameworks developed for several challenges have been given in Table 6.  

\subsection{Object-Oriented Approach }

The object-oriented approach is commonly used for system analysis and design, and this has been proven over the years to be one of the most effective ways of developing systems to meet the functional requirements of users. Additionally, since security assurance involves significant evidence of the incorporation of security measures in a system, the usage of the Demster-Shafer was used to provide such related trust. Demster-Shafer\cite{beynon2000dempster} is a theory of believe functions that is used together with probability and imprecise theories of probabilities for assessing uncertainties. The combined approach in the evaluation model makes the evaluation process of security assurance clearer and makes the results to be more believable.

  Cuihua and Jiajun\cite{cuihua2009object} proposed an object-oriented security evaluation model using the concept of object-oriented technology. The study was motivated by the lack of effective and efficient security models. Existing evaluation models such as qualitative, quantitative, or combined approaches have not provided the necessary security evaluation model. For instance, the quantitative approach cannot obtain precise numerical results. Similarly, the qualitative approach is not objective, and both qualitative and quantitative approach requires significant improvement. To this end, the object-oriented approach and Demster-Shafer evidence were explored for the security evaluation model by Cuihua and Jiajun\cite{cuihua2009object}. They developed a security level distinguishing model using Dempster-Shafer evidence theory to combine experts’ and auto evaluations, thereby making the evaluation process clearer and reliable.

\subsection{Security Assurance Development Process Model }
   Security is an important feature that needs to be considered during the software development lifecycle (SDLC).  A secure SDLC involves integrating security testing and assurance methodology and related activities in the development process. It includes different activities such as security requirements, architectural risk analysis during the design phase, threat modelling, etc. Khan and Khan \cite{khan2018preliminary} made an effort to study the state-of-the-art of security issues, challenges, and practices during SDLC in the industries.  Based on this study, they developed a software security assurance model to assist software developer’s vendor organizations in measuring their readiness to develop secure software.
    
The standard SDLC, OWASP's Comprehensive, Lightweight Application Security Process (CLASP), and Microsoft's security enhancement software development lifecycle (MDL) are not fit for fast-changing commercial requirements to include security assurance measures in operating systems. As a result,  Lan and Han \cite{lan2015sadp} developed security assurance development process (SADP) model.  The model consists of an institute security awareness program, documentation of security-relevant requirements, application of security principles to design, performing source code level review, implementing and performing security tests, security review, and risk management. The activities of SADP can be repeatedly implemented. SADP is proven to be efficient and effective and was hence adopted by NeoKylin operating system.

  \subsection{Vulnerability Detection/Prediction Approach}
  
Vulnerability is a weakness in an information software, internal controls, security procedures, or implementation that can lead to unauthorized access when exploited. Security vulnerability can cause substantial economic and reputational damage to both users and organizations. In the context of software development, vulnerabilities can be introduced in the developed system. Appropriate assurance measures are required to be taken during the development process to avoid the related risks. Therefore, identifying security vulnerabilities in the initial phases of the software development steps is paramount. However, detecting software vulnerabilities is a pretty complex process. 

In the past, some efforts have been made toward identifying and predicting the vulnerabilities, such as Akram and Luo \cite{akram2021sqvdt} developed a quantitative vulnerability detection technique at the source code level based on code clone detection technique. They retrieved vulnerable source code files from the various web source code repositories by tracking the patch file of vulnerabilities. Then, using common vulnerabilities and exposures (CVE) numbers, the vulnerable source code files are retrieved. Dissanayaka et al. \cite{dissanayaka2020security} developed a vulnerability analysis testbed using Linux containers. The authors discussed that the developed virtual testbed is portable and easy to deploy.  Hovseopyan et al.\cite{hovsepyan2016newer} conducted a study by comparing the vulnerability predictions using old and new versions of Chrome and Firefox to determine whether it is better to rely on older versions or newer versions in exploring the vulnerabilities. The study findings suggest that the vulnerability prediction based on the older versions of software tends to be reliable and establish security assurance in that regard. Vulnerability prediction models are one of the possible methods to identify the location of source code that need more attention \cite{hovsepyan2016newer}. 

    In the literature, machine learning methods are also applied to code complexities, code churn, token frequency, developer activities, etc., to detect vulnerabilities towards enhancing security assurance.  Bilgin et al. \cite{bilgin2020vulnerability} used the machine learning technique to predict the vulnerability of the software from source code before its release.  This work also includes developing a source code representation method, intelligently analyzing the abstract syntax tree (AST)  form of the source code, and then verifying whether ML can be applied to distinguish between vulnerable and non-vulnerable code fragments. The vulnerability detection and prediction methods in the development process and on the deployed systems is shown in Fig. 12. 

\begin{figure}
        \centering
        \includegraphics[scale=0.12]{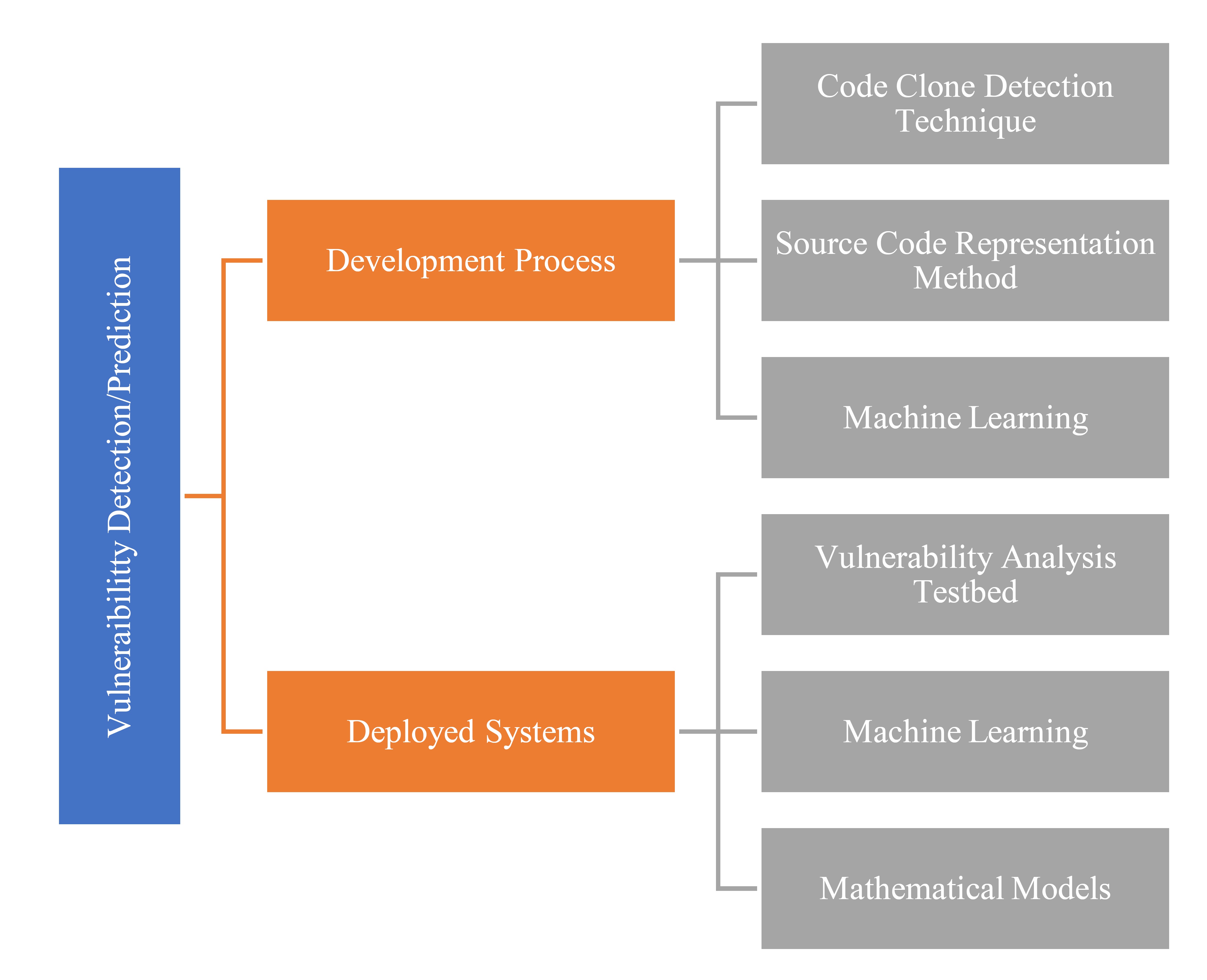}
        \caption{Vulnerability detection/prediction methods.}
        \label{fig:my_label}
 \end{figure}

\subsection{Model-Driven Cyber Range Training}

The increasing demand for security in organizations emphasized the need for security-aware employees in every role of an organization. Therefore, organizations need to train their employees to face potential security challenges and respond accordingly to protect their critical assets. Cyber Range training provides a promising approach that offers employees training in realistic environments considering different scenarios and hands-on experience based on their responsibilities and expertise level. Cybersecurity training should be designed based on the organization's requirements and should adjust to changes in environments and scenarios.  Somarakis et al. \cite{somarakis2019model} studied the importance of dynamic and up-to-date training for cyber security. They proposed a model-driven approach for Cyber Range training, which helps togenerate tailor-made training scenarios based on the organization’s requirements and security posture. This approach also provides the automated deployment of training environments. 

\begin{small}
\scriptsize
\begin{longtable}{|p{.02\textwidth}|p{.12\columnwidth}|p{.28\textwidth}|p{.28\textwidth}|p{.10\textwidth}|p{.04\textwidth}|} 
\caption{Security assurance methods}
\\ \hline
S.N.                       & Security Assurance
  Methods                                   & Challenges Considered                                                                                                                                & Contribution                                                                                                                                                                                                 & Application Domain                                                          & Paper                                            \\ 
\hline
1.                  & Object-Oriented Approach                                     & The existing quantitative
  and qualitative security methods are not efficient.                                                                         & Proposed an object-oriented
  model for a clear security evaluation process structure.                                                                                                                       & Information
  Systems                                                         & \cite{cuihua2009object}        \\ 
\hline
2.  & \multirow{2}{1.5cm}{ Security  Assurance
  Model }                    & Development of secure
  software                                                                                                                        & Developed software security
  assurance model to assist vendor organizations in measuring their readiness
  towards the development of secure software                                                       & Software
  Development                                                        & \cite{khan2018preliminary}        \\ 
\cline{3-6}
                           &                                                                & Current security
  assurance methods are not capable to evaluate the entire development process and
  not applicable for different projects            & Developed an easy-to-implement and
  easy-to-evaluate Security Assurance Development Process (SADP)                                                                                                         & Software
  Development                                                        & \cite{lan2015sadp}              \\ 
\hline
3.  & \multirow{3}{1.5cm}{  Vulnerability 
  Detection/ Prediction  Approach} & Auditing source code
  for vulnerabilities at a large scale                                                                                             &{Developed a
  vulnerability detection method to detect vulnerabilities in software and
  shared libraries at the source code level}                                 & Software                                                                      & \cite{akram2021sqvdt}           \\ 
\cline{3-3}\arrayrulecolor{black}\cline{4-4}\arrayrulecolor{black}\cline{5-6}
                           &                                                                & A model or testbed for
  assessing vulnerabilities that is deployable, maintainable, and accurate                                                       & Developed a general
  vulnerability assessment testbed that is portable, virtual, and deployable                                                                                                             & Software system                                                              & \cite{dissanayaka2020security}  \\ 
\cline{3-6}
                           &                                                                & To predict software
  vulnerabilities from source code                                                                                                  & Developed a source code
  representation method and applied machine learning methods to identify
  vulnerable code fragments                                                                                & Software                                                                      & \cite{bilgin2020vulnerability}  \\ 
\hline
4.            & Cyber Range Training                                           & A cyber security
  program should be tailored to the needs of the organization and should be
  able to adapt to the rapidly changing environment easily & Created a model-driven
  approach for Cyber Range training based on a comprehensive description of the
  organization's security posture that facilitates the generation of tailor-made
  training scenarios & Multi domains including
  smart home, healthcare, smart shipping environments & \cite{somarakis2019model}       \\
\hline
5.  & \multirow{3}{1.5cm}{Security Assurance
  Tools}                       & The need for a tool to
  determine a system's security assurance level                               & Provides a tool for determining
  the level of security assurance of network systems in near real time                                                                                                 & Software
  systems              & \cite{pham2007security,
  pham2008near}  \\ 
\cline{3-6}
                           &                                                                   & Need for a tool to
  assist the evaluator during the CC certification process.                       & Developed a tool to
  represent the CC catalog ontologically                                                                                                                                           & IT
  products and systems       & \cite{ekclhart2007ontological}           \\ 
\cline{3-6}
                           &                                                                   & The need for an
  efficient tool to determine a system security                                      & Created a visualization
  and modeling tool for managing security metrics and measurements in software-intensive
  systems                                                                             & Software -intensive systems     & \cite{savola2011visualization}           \\ 
\hline
6.                  & Attack detection Method & To improve the speed
  and accuracy of attack detection                                              & Implemented machine
  learning and deep learning techniques on the core of the operating system                                                                                                        & Manufacturing
  industry        & \cite{sakthivel2020core}                 \\ 
\cline{1-1}\arrayrulecolor{black}\cline{2-2}\arrayrulecolor{black}\cline{3-6}
7.                 & {Security assurance
  monitoring}                                   & Continuous monitoring
  of telecommunication services is paramount                                   & Demonstrated the applicability of a security assurance methodology for telecommunication infrastructure (BUGYO) on a VoIP infrastructure.                    & Telecommunication service      & \cite{ouedraogo2008deployment}           \\ 
\hline
8.               & Designing Security
  Assurance Framework                          & Assuring the security,
  privacy, and safety of the handled data.                                    & Conceptually aligning
  EBIOS, Secure Tropos, and PRIS methods to create a complete assurance
  framework                                                                                              & Intelligent
  Transport Systems & \cite{diamantopoulou2020aligning}        \\ 
\hline
9.                & Anomaly detection method                                         & Monitoring and
  analyzing network security to identify suspicious activity throughout the
  network & Discussed the available methods and tools and presented
  a general concept of security analysis component for the security operational
  center.                                                    & Network
  Traffic               & \cite{bialas2019anomaly}                 \\ 
\hline
10.                 & Assurance case method                                             & The adaptation of
  existing assurance case methodologies to cover specific IIoT issues              & Presented a framework
  for assurance cases for IIoT and security cases that included elements of
  properties assurance, security management assurance, component assurance, and
  feature assurance. & Industrial
  IoT                & \cite{sklyar2017challenges}              \\
\hline
11.                 & Continuous assurance
  methods                            & To assess the security
  of IoT systems continuously, a systematic approach is required & Developed a conceptual
  framework for IoT security assurance evaluation and presented a process for
  developing continuous assurance methods for IoT services.                                                                                                           & IoT                    & \cite{ardagna2018case}              \\ 
\hline
12.                  & Model-Driven Security
  Assurance Framework               & Security assurance
  needs to be automated                                              & Proposed a model-driven
  framework to allow designers to model security concerns and to facilitate
  automated verifications and validations.                                                                                                                             & Open-source
  software & \cite{rauf2017towards}              \\ 
\hline
13.                 & Security assessment
  system                              & Secure deployment of
  WLAN network                                                     & Developed a fuzzy
  logic-based security assessment system                                                                                                                                                                                                                 & WLANs                  & \cite{liu2015saew}                  \\ 
\hline
14.  & \multirow{2}{1.5cm}{{E-government security\\
  assurance system}} & Constructing safe,
  reliable and effective e-government security assurance             & Developed a reference
  model for the security assurance of e-government systems                                                                                                                                                                                           & E-governance           & 
  \cite{xiang2010analysis}  \\ 
\cline{3-6}
                           &                                                           & To educate personnel working in E-Government
  systems about cyber security assessment & Created a framework for
  assessing information security issues in e-government                                                                                                                                                                                            & E-governance           & \cite{gupta2020cyber}               \\ 
\hline
15.                 & Security Protection Profile                              & Railway automation does
  not have any harmonized IT security requirements              & Describes a reference
  communication architecture which attempts to separate risk management and
  security requirements as well as certification processes as far as possible, and
  outlines the threats and IT security objectives with typical railway
  assumptions. & Railway
  Automation   & \cite{bock2012towards}            \\
\hline
\end{longtable}
\end{small}
\subsection{Usable-Security Assessment}

      The usable security software service may be subject to different security attributes such as confidentiality, integrity, availability, accountability and non-repudiation, and usability attributes such as effectiveness, efficiency, satisfaction and user error protection. These attributes are crucial in the security assurance of the software systems and will help to increase the efficiency and accuracy in security assurance measurement.  Agrawal et al. \cite{agrawal2019multi} used the Fuzzy Analytical Hierarchy Process (Fuzzy-AHP) methodology to evaluate usable security. They also assessed the impact of security on usability and the impact of usability on security using a quantitative approach.

\subsection{Security Assurance Tools} 

Security assurance tool supports measurement of real-time security assurance level and to maintain the security of complex and large-networked operational systems in automated way. Pham et al. \cite{pham2007security, pham2008near} discussed the need of security assurance tool to measure the assurance level of the deployed system. Some authors also made efforts towards the development of tool to support security assurance evaluation process.  

CC certification is a complex and time taken process that demotivates industries to adopt this process. Ekclhart et al. \cite{ekclhart2007ontological} developed CC ontology tool to ease the CC certification process. This tool is based on ontological representation of CC catalog and supports various tasks such as evaluation process planning, making reports and review of the applicable documents. Savola and Heinonen \cite{savola2011visualization} developed a tool for visualization and modeling the hierarchical specification and deployment of security metrics and measurements. This tool also supports security decision making by managing a large number of metrics and measurements in an efficient way.  
     
\subsection{CC Ontology Tool}
  The certification process of information security products and systems using CC is very complex and time-consuming. This has discouraged several organizations from the CC certification process. To address this issue, Ekclhart et al.\cite{ekclhart2007ontological} developed a CC ontology tool based on an ontological representation of the CC catalogue. The developed tool supports the planning of an evaluation process, reviewing the relevant documents, and creating the reports. These features simplify the CC certification process and can facilitate the adoption of CC certification by more organizations. However, the developed tool is mainly designed for security assurance requirements relevant for the evaluation.

\section{System and Environments}

The recent advancements in security assurance suggest a promising approach to ensure the security of software. The security assurance of the system depends on both static and dynamic security metrics. The static metrics of the system change when there are changes in configuration or a component is added or removed. Whereas the dynamic metrics change over time \cite{pham2008near}. However, most of the existing works focus on the static vision on the description of the system under evaluation and do not consider the system's dynamic behaviour. For example, the CC is a static process that relies on the accredited security experts. The main disadvantage of this procedure is that it is ad-hoc in the laboratory for a single product, and it loses significance when a certified product is brought into an operational system or modified.  Implementation of operational security assurance in large dynamic systems is a challenging task. Moreover, providing continuous security assurance evaluation and implementing countermeasures to achieve the security goals is also a complex task \cite{haddad2011operational}.    Fig. 13 represents the evolution of static and dynamic security assurance evaluation methods.  This figure shows that most research works have considered static security assurance, whereas very little research has been done on evaluation of dynamic security.

 \begin{figure}
        \centering
        \includegraphics[scale=0.45]{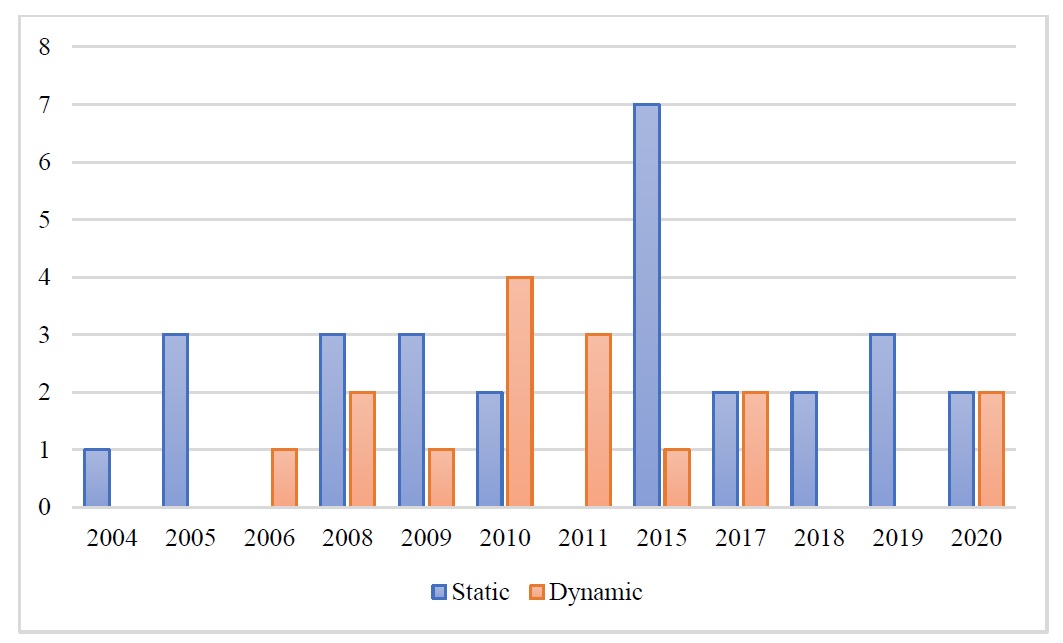}
        \caption{Static and dynamic security assurance methods.}
        \label{fig:my_label}
      \end{figure}
    
    The development of the security assurance method of a software system needed the definition and a clear understanding of the security requirements of the concerned application domain. Many researchers proposed the general security assurance methodology, while others developed the application-oriented approaches. As per our observation, in some application domains, pervasive research works have been done, such as cloud computing, telecommunication, etc. while in some of them, there are very limited research works such as railway, e-governance, etc. The overview of the application domains in which the research works related to security assurance have been shown in Fig. 14. The application domains are categorized mainly into two categories: CPS and IT systems. Since cloud is quite a popular application domain and many research works have been done. Therefore, we will discuss the challenges and methods of cloud security assurance separately. 
\begin{figure*}[hbt!]
        \centering
        \includegraphics[width=14cm,height=12cm]{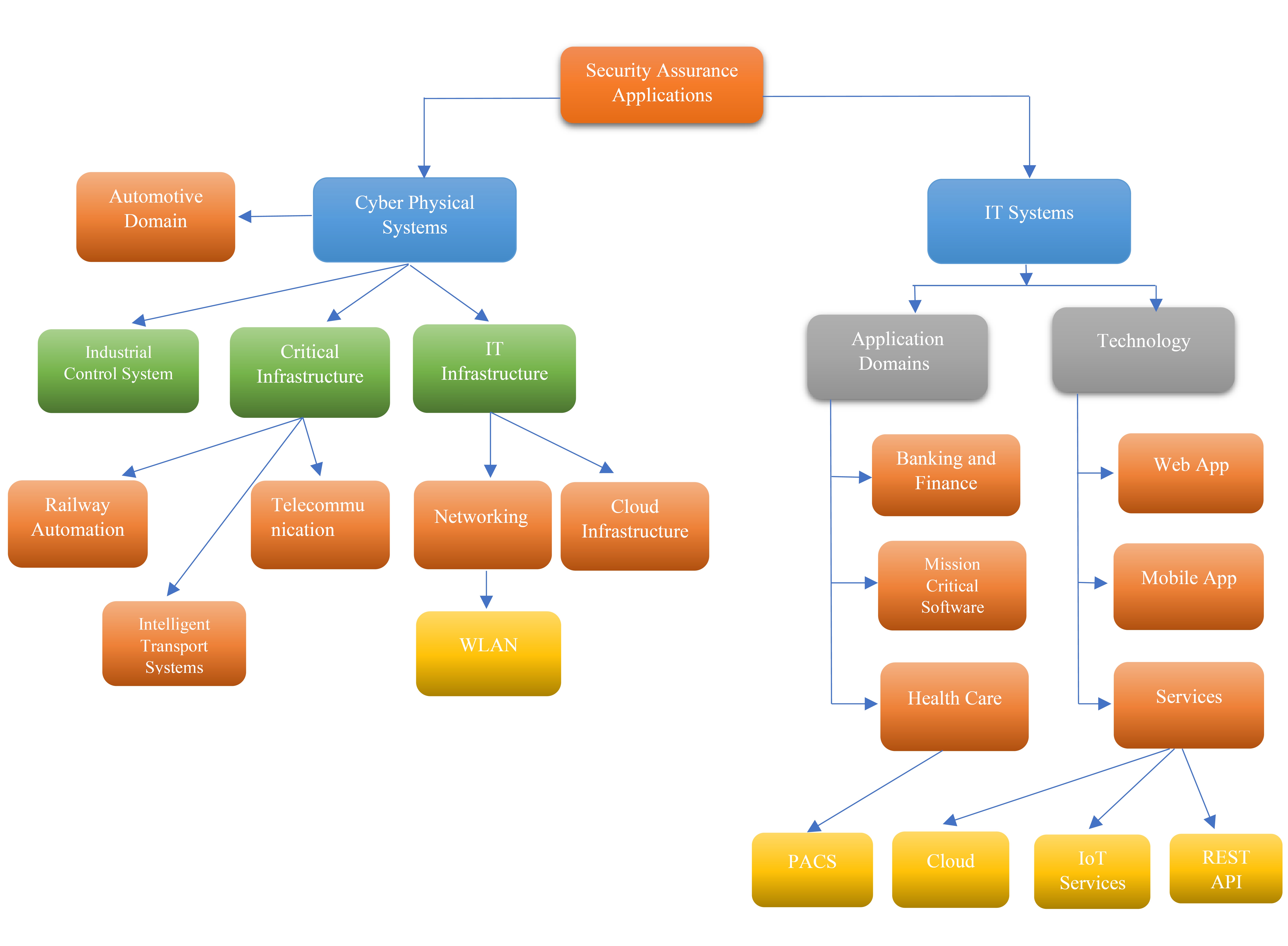}
        \caption{Application domains of system security assurance.}
        \label{fig:my_label}
    \end{figure*}
    
 \subsection{Security Assurance in Access Control }
Access control is one of the important measures in security assurance and crucial to ensure that information is secured, uncorrupted, and available in the application. A system should have the assurance that the resources access by the users are authorized and authenticated. The different types of access control include role-based and attribute-based \cite{nweke2020understanding}. Therefore, the incorporation of access control needs to be considered in the initial stages of the system development process. 

     It can be a challenging task to incorporate the access control into the software if most of the requirements related to the access control are identified and implemented after the functional requirements \cite{devanbu2000software}. Accordingly, access control requirements are mostly considered after the functional requirement definition stage, leading to flaws and security defects in the access control mechanism. Therefore, a security assurance mechanism is required to ensure that the application code behaves consistently with the access control policy \cite{pavlich2010framework}. A framework developed by Pavlich-Mariscal et al. \cite{pavlich2010framework} showed that the access control requirements that were identified and designed have been correctly implemented at the code level. A labelled transition system (LTS) was used at the design stage to represent the behaviour of the system. Lambda-calculus was used at the code level to capture the essential features of the access control requirements.  The requirements were then mapped from the LTS to the lambda-calculus as a representation of the correct representation of design to the code. The work was validated by applying the framework to two access control mechanisms.

\subsection{Security Assurance of the Self-Adaptive System }
The trustworthiness of the self-adaptive system is the primary concern. The dynamic environment of these systems makes security goals challenging to achieve. These systems must manage uncertainty due to the dynamic environments, functional changes and malicious attacks and fulfil their security requirements in the deployed environment and as if there are any changes over time. Therefore, security assurance of these systems is required to maintain the system's trustworthiness within the uncertainty.  The security assurance of these systems becomes even more critical when they are used in business-critical and safety-critical applications.  Therefore, an efficient method is required to assure confidence in the security profile of self-adaptive systems, which adopts the changes in functional and security conditions at runtime. However, it is a challenging task due to the complexity of the security dependencies.   
To manage the security concern, the self-adaptive system should consider MAPE (Monitor, Analyse, Plan, and Execute) control loop \cite{le2015quality}.  In this regard, Jahan et al. \cite{jahan2020mape} developed a security-focused feedback control loop, MAPE-SAC, and its interaction with a MAPE-K, function, and performance-focused control loop to dynamically manage runtime adaptations in response to changes in functional and security conditions.

 \subsection{Core-level cybersecurity assurance}

Sakthivel et al. \cite{sakthivel2020core} studied the core-level cybersecurity assurance. They considered the core of the operating system than the application level service for their study to increase the speed and effectiveness in attack detection.  The main advantage of considering the operating system's core is that it contains every internal attribute and the file system. They implemented the machine learning and deep learning approach and found that classification and learning methods used in different machine learning techniques can enhance the protection of the systems against potential attacks.

\subsection{Security Assurance of Telecommunication System}
Telecommunication infrastructure experiences continuous and increasing security threats and risks. There is a need for the deployment of adequately tested and managed information security solutions. The security assurance of the telecommunication system can help to fulfil this requirement. Software security assurance incorporates different methods such as security testing, security analysis, security monitoring, security auditing, etc. Security testing is an essential process in security assurance, and it should be implemented iteratively on analytical and practical stages \cite{savola2009software}. On the other hand, continuous monitoring of telecommunication services is paramount to get the quality assurance of the services being provided. Ouedraogo et al. \cite{ouedraogo2008deployment} discussed the applicability of the BUGYO (Building Security Assurance in Open Infrastructures) methodology on VoIP service infrastructure based on open source components. BUGYO is a CC-based methodology that addresses the security assurance issues related to the telecommunication infrastructure and services and verifies that the services provided through the infrastructure are secured. 

\subsection{Security Durability of Web Application}
Kumar et al. \cite{kumar2020knowledge} focused on optimizing the security assurance effort for a specific lifespan of a web application by estimating its security durability. They identified three factors: trustworthiness, dependability, and human trust, which can improve security durability. They estimated the security durability quantitatively by using a Hesitant Fuzzy based Analytic Hierarchy Process and Technique for Order of Preference by Similarity to Ideal Solution hybrid technique.

\subsection{Security Assurance Framework for Intelligent Transport System}
The growing number of vehicles on the road is one of the significant causes of road traffic accidents, a major public safety problem. The emerging model of connected vehicles enables a significant technology shift in improving road safety and fostering the emergence of next-generation cooperative  intelligent transport systems (ITS). Networked ICT, internet of things (IoT), and CPS help the ITS enable services and applications such as communication of vehicles with nearby vehicles and roadside infrastructures, etc.  However, these technologies come with several challenges, including security and privacy.   The wireless connection of these vehicles to external entities can expose the ITS applications to various security threats and attacks. Therefore, it is essential to consider these concerns while developing the ITS.  Diamantopoulou et al. \cite{diamantopoulou2020aligning} developed a security assurance framework for connected vehicular technology.

\subsection{Security Assurance of Network Traffic}
With the rapid growth of organization intranets, network traffic security has become a more complex task for network security administrators. The incorporation of emerging technologies such as cloud computing and IoT is also increasing the attack surface.  Therefore, it is vital to monitor and analyze the network security to identify suspicious activities throughout the network and mitigate it. The security analysis process includes the intercepting, recording, and analysis of the pattern of network traffic communication. The implementation of security measures in a particular organization is incomplete without these processes. Bialas \cite{bialas2019anomaly} considered one of the main concerns of network security assurance, i.e., anomaly detection in network traffic monitoring and developed an anomaly detection component. 

\subsection{Security Assurance of IoT Devices}
Present-day, IoT has become one of the unprecedented research topics in cybersecurity research. IoT is a system that enables computing devices, mechanical and digital machines, objects, things, and animals in our environment to interconnect with each other over the internet without the need for human-to-human or human-to-computer interaction. Embedded sensors and unique identifiers allow smart things to interact with each other over the internet; and create smart applications and services such as smart cities, smart homes, smart schools, smart healthcare smart cars, intelligent transportation, smart grid, etc. IoT provides real work intelligent platform for distributed object interaction through various communication technologies. It made IoT systems an attractive target for attackers and adversaries interested in stealing sensitive information. Therefore, a systematic method is needed to assess the security of IoT systems.   Ardagnan et al. \cite{ardagna2018case} developed continuous assurance methods for IoT services and designed a conceptual framework for IoT security assurance assessment. Sklyar and Kharchenko \cite{sklyar2017challenges} proposed a framework to utilize Assurance Case methodology for Industrial IoT systems (IIoT).

 \subsection{ Security Assurance Framework for OSS} 
 Open source technology is a growing trend in a wide range of applications. There is a great demand for open source technology in most modern enterprises.   However, it comes with great concern over the security assurance provided by open source components. Frequent updates are required to fine-tune the product and to improve its performance. These updates add new features and improve or remove old features in the software; this can violate the security properties designed for the previous versions.    Therefore, a systematic and efficient security assurance technique is required to measure the security level of open sources software and its released versions in an automated way.  When open source software (OSS) is combined with the REST architectural style, this task becomes more challenging.  REST architecture has several additional scalability and extensibility benefits,  which help to add more features and offer the services to a larger audience. Design methodologies and mechanisms are required to use the REST APIs that manage stateless protocol for stateful applications. Rauf and Troubitsyna \cite{rauf2017towards} proposed a model-driven framework that will help designers to model the security concerns of REST compliant open-source software and to enable automated verification and validation. It also facilitates a regular monitoring mechanism of the security features, even considering frequent updates of OSS.  
 
 \subsection{Wireless Local Area Networks (WLANs)}
 
Wireless networking is quite popular and one of the rapid growing technologies. A WLAN is a network that enables devices’ capability to connect and communicate wirelessly for homes and businesses.  However, while it provides convenience and flexibility, secure deployment is not always possible and is still a primary concern for researchers and developers.   Liu and Jin \cite{liu2015saew} conducted a study to analyze the security threats and attacks to the WLAN network architecture and developed a security assessment and enhancement system. This system is divided into two subsystems, a security assessment system and a security enhancement system.  The security assessment system is based on fuzzy logic and analyzes vulnerability of physical layer (PHY) and medium access control (MAC) layer, key management layer and identity authentication layer. This approach provides quantitative value of security level based on security indexes.  Whereas the security enhancement system is an integrated, trusted WLAN framework based on the trusted network connect that helps to improve the security level of WLAN.

 \subsection{Security Assurance of E-governance}
 
 E-government can be defined as the use of information and communication technologies in delivering government services across the citizens, various public service agencies and businesses, and the management of these services effectively and efficiently. E-government helps the government to reduce operational costs by better-integrated workflows and processes and effective utilization of resources. However, e-governance comes with several challenges, among which information security is the more serious. Therefore, the government should take the necessary measures and actions to secure useful information such as government data and citizens' data to guarantee national security. The security assurance of E-governance can guarantee the operation of the e-government system in a secure, reliable, and effective way. Lixiang \cite{xiang2010analysis} discussed the security issue of Chinese E-governance and proposed a reference model for information security assurance. Gupta et al. \cite{gupta2020cyber} conducted a study to educate the person working in the E-governance system. In this regard, they suggested a framework considering various internal and external features of an organization's security.  

\subsection{Protection Profile in Railway Automation}

Railway automation has experienced increasing IT security incidents in recent years. However, most of these attacks are denial of service attacks to target interruptions of services, not related to any safety-critical incidents.  Therefore, IT security requirements for railway automation is needed. Bock et al. \cite{bock2012towards} discussed this issue and represented the CC-based IT security requirements for the railway domain as a protection profile. They also developed a reference architecture to distinguish IT security and safety requirements and discussed security objectives and threats in this application domain.

\subsection{Cloud Security Assurance }
Cloud computing is a computing paradigm that has transformed the service deployment landscape of ICT. It offers various on-demand, highly available, scalable, and ubiquitous services that extend beyond the geographical and administrative boundaries, thereby attracting organizations to deploy their services in the cloud.  However, cloud services come with security concerns due to many users. The externalization process of the application is also one of the main reasons to increase its potential surface of attack. The lingering concern of security concerns is one of the main reasons organizations resist the cloud.  Thus, cloud security became the major concern for both researchers and industries. The challenges face by CSCs, CSPs and common issues are illustrated in Fig. 15. The overview of assurance methods and techniques developed in the literature are as follows:  
\begin{figure*}[hbt!]
        \centering
        \includegraphics[scale=0.47]{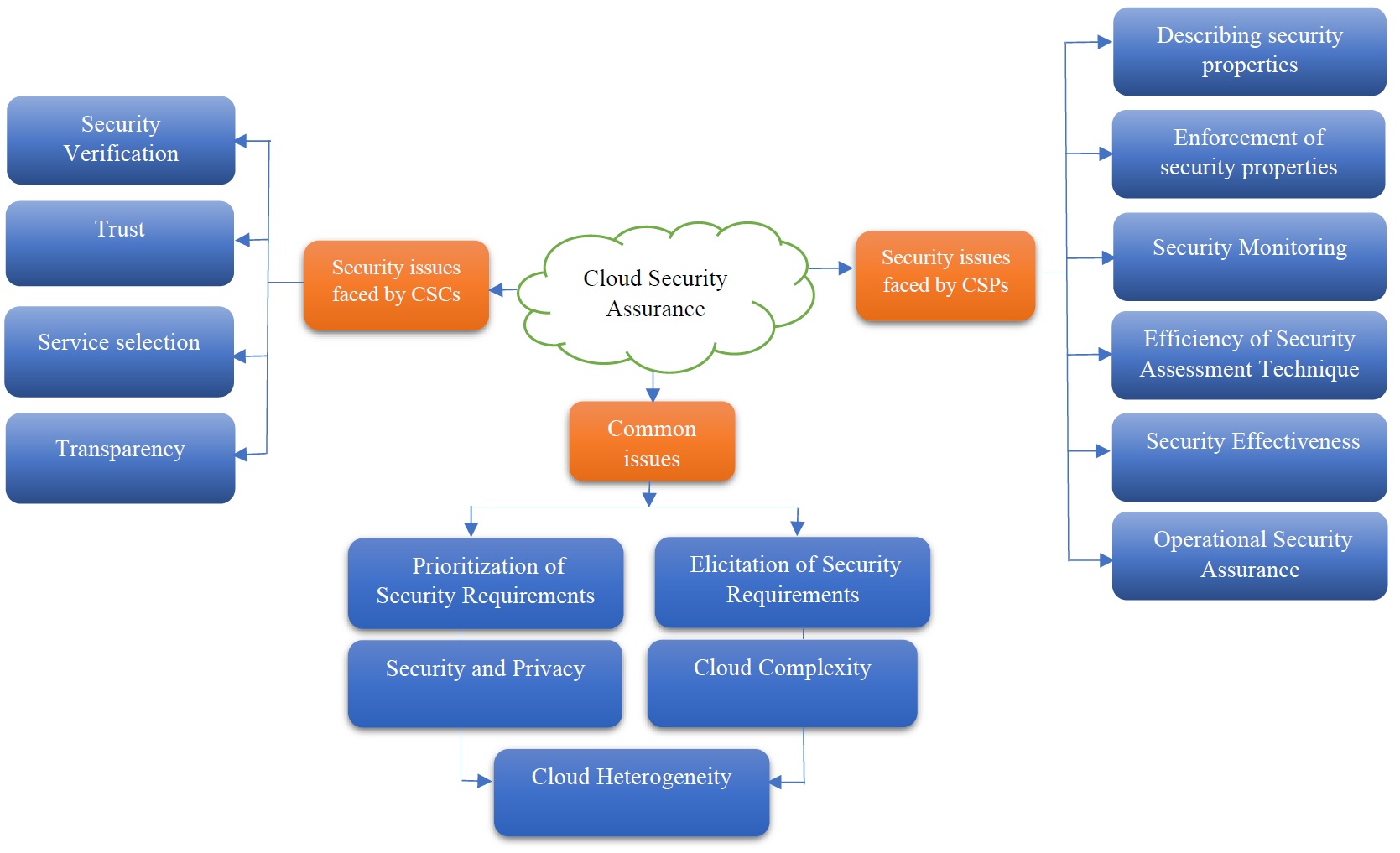}
        \caption{Security issues related to cloud security assurance.}
        \label{fig:my_label}
    \end{figure*}
\subsubsection{	Enforcement of Security Properties in Cloud }
    
 Enforcement of the security properties in the cloud is an essential and challenging task. It is important to assure that the essential security requirements have been effectively enforced. The assurance properties can be helpful to verify the effective enforcement of the security requirements and provide evidence of the enforcement. Bobelin et al. \cite{bobelin2015autonomic} proposed a context-based language to express the assurance properties based on the security requirements. The property prototypes are used to represent the security requirements. These prototypes are high-level definitions of security requirements that can address one of the major issues, i.e., difficulty in defining the security policy.

\subsubsection{	Security Assessment of Cloud Considering CSPs and CSCs Requirements}
    Cloud computing provides multiple benefits to users. However, CSPs should provide trust to CSCs with their data and ensure data integrity and confidentiality.  On the other hand, CSCs should know what security assurance CSPs are offering and should compare the CSPs offers qualitatively or qualitatively based on their requirements.  The formalization of the security properties CSPs offer is the first step in assessing and comparing CSPs.  Cloud community is working to develop a standard specification of security service level agreements (SecSLAs) that can help CSPs to ensure the desired level of security to CSCs for their services \cite{hubballi2019cloud}. 
    
 Modic et al. \cite{modic2016novel} developed a cloud security assessment technique, i.e., moving intervals process. This method supports multiple requests in parallel, and it also considers variation in the number of CSPs and the security requirements of CSCs.  The proposed technique offers high computational efficiency as well as accuracy. They also developed an approach to quantify the security attributes considered in SecSLAs to compare different SecSLAs offered by CSPs.  Rizvi et al. \cite{rizvi2018security} also considered the decision-making problem of CSCs in the selection of available cloud services based on their security strength. They proposed a framework and mechanism that can assess the robustness of CSP security based on customer security preferences. They developed security evaluation rules for security metric based on linear equations. This rule helps cloud service users analyze the security index scores of one or more CSPs and verify the final security scores.

\subsection{Continuous Security Assessment for Security Certification of Cloud}
     Cloud services need continuous refinements and requirements changes, which affects the security and resilience posture of the service and invalidates the certificate.  In this situation, recertificate may be needed, which leads to high costs. These drawbacks demotivate the CSPs that intend to leverage the advantages of the cloud.
     
 In the last decades, several methods have been developed for systematic and comprehensive cloud security to address this shortcoming.  One of the drawbacks of these approaches is that they can reveal a cloud service provider’s intellectual property as part of the assessment process. Aggregation of assessment results into the assurance levels can be one of the possible methods to address this problem, which is used in CC \cite{herrmann2002using}. CC assurance levels show the sophistication level of tests performed on a system, such as functional, structural, methodical, semi-formal, and formal. Hudic et al.\cite{hudic2017security} made an effort to develop a security assessment methodology for security certification, which provides continuous security assessment and also protects the cloud provider's intellectual property.

\subsubsection{ Security Transparency and Audit}
Security transparency and audit are essential factors that the industries must consider to increase the users' trust by fulfilling the requirement of the offering services. Security transparency increases assurance and accountability by providing essential information to the customers about the security practices and procedures. Whereas, security auditing is the tracking and collecting evidence of the significant events in the operational phase of the system, which is useful to achieve the overall goal security objectives of the system.  Ismail and Islam \cite{ismail2020unified} discussed these two factors and developed a framework to address challenges regarding security transparency of the cloud. They have also developed a Security transparency and audit tool that can help auditors to evaluate the evidence produced by the CSP.

\subsubsection{ Enforcement and Monitoring the SecSLAs} 

The establishment of SecSLAs is not enough to fulfil the requirements of CSPs and CSCs when there is a lack of management in SecSLAs commitments. A systematic mechanism is required for CSPs to take action against the eventualities such as attacks, changes, disasters, or regulations that may affect the fulfilment of the SecSLA commitment. The efficient detection and mitigation of potential threats or harmful security-related events are essential for CSPs to provide trustworthy services to CSCs and fulfil and maintain the agreed assurance levels. For example, if an attacker cracks the encryption key, there is no automated mechanism to indicate that there may be a security breach and the protected information is exposed. Trapeiro et al. \cite{trapero2017novel} proposed a solution for monitoring and enforcing SecSLA compliance.

A summary of cloud security assurance methods has been given in Table 8.

\begin{table}
\caption{A summary of cloud security assurance methods}
\scriptsize
\centering
\begin{tabular}{|p{.03\textwidth}|p{.26\textwidth}|p{.26\textwidth}|p{.28\textwidth}|p{0.04\textwidth}|} 
\hline
S.N.                        & Category                                                & Challenges                                                                                                                                   & Proposed Solution                                                                                                                                                                     & Paper                                         \\ 
\hline
1. & Security
  assurance of cloud services & Standard
  specification of security service level agreements                                                               & - Developed a cloud security assessment technique 
  
  -Developed an approach to quantify the security attributes considered in SecSLAs         & \cite{modic2016novel}        \\ 
\cline{4-5}
                            &                                                         &                                                                                                                                              & Proposed
  a framework and mechanism that can assess the robustness of CSP security
  based on customer security preferences                                                          & \cite{rizvi2018security}     \\ 
\hline
2.            & Enforcement
  of the security properties                & Effective enforcement
  of security properties in the cloud                                                                                  & Proposed a
  context-based language to express the assurance properties based on the
  security requirements                                                                          & \cite{bobelin2015autonomic}  \\ 
\hline
3.            & Security
  Certification of Cloud                       & Continuous
  refinements and requirements changes may invalidate the cloud certificate                                                       & Develop a security
  assessment methodology for security certification                                                                                                                & \cite{hudic2017security}     \\ 
\hline
4.            & Security
  transparency and audit                       & Implementing
  security transparency and audit                                                                                             & -Developed a framework to address challenges regarding security transparency of the cloud 
  
  -Developed a security transparency and audit tool & \cite{ismail2020unified}     \\ 
\hline
5.            & Enforcement
  and Monitoring the SecSLAs                & A
  systematic mechanism is required for CSPs to take action against the
  eventualities such as attacks, changes, disasters, or regulations & Proposed a solution for
  monitoring and enforcing SecSLA compliance.                                                                                                                 & \cite{trapero2017novel}      \\
\hline
\end{tabular}
\end{table}

\section{Limitations and Future Research Directions}
   
The rapid evolution of technology comes with security concerns and lends the information technology environment to potential uncertainty. Therefore, security solutions must be advanced to prevent potential threats and risks before and when they occur.  The current security assurance methods are good at flagging the anomalies but not too good at defining their impact and potential risks. These are some limitations of the existing security solutions and future challenges:

\subsection{Security Requirements Analysis}
CC provides organized and concrete guidelines for evaluation. CC sets the possible security requirements that include assurance requirements and functional requirements. However, the elucidation of the structural relationship between security requirements and security assurance requirements and their evaluation is a possible future research direction \cite{taguchi2010aligning}. 

\subsection{Enforcement of Security Requirements}  
The security requirements are developed to meet the need of consumers, developers, and evaluators. To achieve a particular EAL, a set of activities including all security requirements needs to be identified and evaluate these security requirements against the user’s requirement and eliciting new requirements based on the objectives to be met.  However, CC does not offer a well-defined methodology to employ the enforced requirements. 

\subsection{Correctness, Quality and Effectiveness of the Security Assurance }
Security assurance provides confidence that system assets are protected; however, this confidence depends on the correctness of the security measures, quality of the security policy, and profile of the attackers \cite{ouedraogo2010agent}. The effectiveness of the assurance techniques is another factor that is difficult to measure in the operational phase. A methodology is required to measure that the security controls are adequate and appropriate for specific systems. Therefore, it is essential to consider these factors while developing a security assurance method.   

\subsection{Real-time Security Assurance Evaluation} 

In most of the existing works related to security assurance evaluation, the system's dynamic behaviour under evaluation has not been considered. These works mainly focused on static behaviour. Therefore, a method/tool is required to measure and verify the security assurance level of the system in real-time. This method will be helpful to maintain and improve system security on a day-to-day basis in the operational phase.

\subsection{Automation of Security Assurance Process}
Automation of security analysis, evaluation, validation, and testing approaches is required to obtain continuous evidence about a  system's security assurance level or performance under evaluation. There are no widely accepted methods or approaches that fulfil this requirement. 

An automated security verification technique for web applications is also crucial. The current technology, such as network firewalls and antivirus, offers security protections only at host or network levels. Therefore, one should also focus on security at the application level, i.e., the publication interfaces of web applications which is a more attractive target. In the past, white-box and black-box testing frameworks have been used for automated web applications security assessment. However, one of the significant drawbacks of black-box testing is that it requires source code which is not available in many cases, and the second is that the verification process runs on simulated runtime behaviours based on program abstraction, while the abstraction may not reflects the actual program correctly. On the other hand, black-box testing framework work also comes with several challenges, such as providing an efficient interface mechanism \cite{huang2005testing}.

\subsection{Security Assurance Tool}

Security assurance tools help developers in building secure systems and to measure how secure system is.  There is no efficient tool available that supports real-time security assurance of the system in an automated way. Therefore, there a need is to improve the security assurance tools and develop new tools that can help in security assurance evaluation of large networked IT systems. 

\subsection{Security Assurance of Composed System}

Direct measurement of security assurance for complex software systems is not always possible. In this case, one of the possible methods is to decompose the system into measurable parts and measure the assurance of decomposed entities. However, in this method, aggregation is required as a backward process to obtain the security assurance of the software system by aggregating the security assurance values. Therefore, a suitable aggregation technique is required. The method should consider the composing entities that are security assurance relevant and relations between the components. It is also essential to understand how constituent components of a system under evaluation contribute to the system’s security \cite{pham2007security}.  Defining and developing confidence metrics for security assurance level and the selection and aggregation technique for these security metrics is another future direction. 

\subsection{Quantitative Security Assurance Approach }
In the past, very few efforts have been made to develop a systematic approach to quantify the security to support security assurance activities. Most of the approaches are focused on qualitative security assurance.  There are no standards or widely accepted methods, models, taxonomy, or tools for quantitative security assurance.  The advantages of the quantitative security assurance approach have already been discussed in the previous section. 

\subsection{Competencies of individuals }

Individuals performance assessment and verification are essential for different industries and organizations. A possible future direction is to develop an assurance technique that can assess the competencies of the individuals for conducting the security assessment \cite{knowles2015assurance}. This technique will be helpful for an individual in improving security assurance skills.   

\subsection{Development of Metrics Visualization System }

The security metrics measurements and meaningfulness in security assurance and risk management is a challenging task.  Their poor manageability can compromise the meaningfulness of the metrics due to a large number of uncategorized information elements. Therefore, systematic methods and well-designed tools are required for better management in collecting and measuring security metrics. In the past, limited research work has been done in this area, and they are at the initial research stage \cite{savola2011visualization}.  

\subsection{Data Driven Security Assurance Method}
The recent advancements in computing capabilities, software algorithms, and specialized hardware design have led to major breakthroughs in machine learning and artificial intelligence. The application of data-driven techniques in security assurance may provide a promising solution in automated and intelligent security analysis, including vulnerability identification, code classification, vulnerability prediction, code summarizing, and clone detection. In the literature, limited research work can be found in this area \cite{bilgin2020vulnerability, kumar2020knowledge, sakthivel2020core}.

\subsection{Vulnerability Assessment}
In the literature, several methods and techniques are developed to identify and predict the vulnerability. However, certain factors need to be considered while developing these methods such as “how these vulnerabilities transfer from one system to another?” and “how these vulnerabilities remain unfixed for a long time?” \cite{akram2021sqvdt}. Another future direction is identifying the exact location of detected vulnerability in the functional level code and what it is the reason behind its detection as a vulnerability. In other words, considering and improving the localization and interpretation aspects of the vulnerability \cite{bilgin2020vulnerability}.

\subsection{Cloud Security Assurance }

It is a quite difficult task to enforce the security properties on a cloud platform. However, it is a complex task to describing and enforcing security properties and collection of digital evidence back. The multi-layered system and services and their interdependencies make the security evaluation even more difficult. CSPs should assess the security of their offering services. Therefore, there is a need for a method or technique that address these issues. Some other requirements are as follows: 

\subsubsection{Security Monitoring Tool}

The tenants want assurance that the security properties of the requested services are enforced effectively.  Therefore, CSCs should offer a tool or way so that tenants can monitor the actual security of service and can provide well-defined information regarding violations of requested security features.  

\subsubsection{Real-time Cloud Security Assessment }

Some security assessment methodologies have been developed in the literature, but they are worthy only if they can support the actual decision-making at run time. Most of these methods are effective in an environment where performance is not essential \cite{modic2016novel}. Therefore, an efficient method is required which support operational security assurance evaluation.   Future research work should focus on the real-time monitoring of cloud security. These techniques should also include multi-cloud applications and a cloud supply chain. It should be able to adopt the changes in both cloud service composition, and the context \cite{rios2017dynamic}.

\section{Conclusion}
ICT has a significant impact in this information age era. Public and private organizations heavily rely on software systems to make their business activities more manageable.  However, these systems are experiencing increasing threats with the evolution of new technology; hence organizations need to expand security assurance programs to encounter the security threats and ensure their integrity and quality.  Security assurance provides confidence that the system meets the security requirements and is resilient against potential security threats and failures. It involves different processes such as requirement analysis, design, implementation, verification, and testing. 

Traditional security assurance methods suffer from several limitations due to their static nature, poor security requirements specifications and design, etc. These methods are also focused on either a single application or component of a system.  This paper has conducted a comprehensive SLR to study state-of-the-art, research trends and future directions in system security assurance. We have organized and presented detailed discussions on the security assurance  processes and activities, such as assurance criteria, profile, methods, assurance level, system, and environments.  In this paper, we have discussed the research challenges and gaps and different security assurance techniques and methods that have made an effort to overcome these challenges. We have also discussed the limitations of these security assurance methods and future directions.    The primary limitations that current approaches cannot resolve are entirely related to the security assurance specification and enforcement, correctness, effectiveness of security assurance, automation of security assurance process and real-time security assurance evaluations. Therefore, an advanced security assurance processes and methods will be required to ensure an organization's required level of security. 

\section*{Acknowledgment}

This work was carried out during the tenure of an ERCIM ‘Alain Bensoussan’ Fellowship Programme. This work is also partially supported by Norwegian Cyber Range, NTNU, Norway.





 \bibliographystyle{model1-num-names}

\bibliography{cas-refs}





\newpage
\appendix
\begin{appendices}
\section{Keywords for Data Extraction and their Definitions}
\begin{enumerate}
    \item 	\textit{Challenges and Gap}  \item	\textit{Contribution} 
 \item	 \textit{Process}:  It is the series of action or steps taken in order to develop security assurance framework/evaluate the security assurance. 
 \item	 \textit{Methods}: Which methods or techniques used for security assurance evaluation for example mathematical methods/Fuzzy Techniques/Data Driven Approach. 
 \item	\textit{Guidelines/Standards}: Any standards/guidelines used to define the security goals/security requirement/security metrics/assurance measurement, for example ISA/IEC 6151, ISA/IEC 62443, NIST 800-82 R2, ISA-TR84.00.09
 \item	\textit{Tools}: Is there any tools which are developed or used for testing/scanning/security assurance measurement, for example: OWASP Zed Attack Proxy (ZAP), WebScarab, OpenVAS. 
 \item\textit{	Metrics}: Security metrics are used to diagnose issues, identify weak links in the existing security posture, facilitate benchmark comparisons, and derive performance improvement. Security metrics derived from the security objectives. In this SLR, Security Metrics are the metrics that have been considered in security assurance framework, identifying threats, security requirement and risk assessment for example: Confidentiality, Integrity, Availability, Exploitability, Vulnerability, Host Verification, Guest Verification, Reliability, authentication, authorization, etc.  Some of papers may define other security metrics based on their security goals.

\textit{Note}: If the metrics are not mentioned in the general framework of the security assurance then metrics mentioned in the test cases/examples/illustration will be considered in the data sheet and the application domain will also be mentioned accordingly. In case of the more than one application domain, all the application domain and respective security metrics will be mentioned. 
 \item\textit{	Evaluation/Techniques}: Whether it is qualitative or quantitative?
 \item	\textit{Automation}: Whether the vulnerability detection/countermeasures/ testing process or security assurance process are automated/semiautomated or manual?
 \item\textit{	Application domain }
 \item\textit{	Limitations and direction}
 \item\textit{	Number of citations} 

\end{enumerate}

\end{appendices}
\end{document}